\documentclass[12pt, draftclsnofoot, onecolumn]{IEEEtran}
\usepackage{cite}
\usepackage{amsmath,amssymb,amsfonts}
\usepackage{graphicx}
\usepackage{textcomp}
\usepackage{xcolor}
\usepackage{bbm}

\usepackage{algorithm}% http://ctan.org/pkg/algorithm
\usepackage{algpseudocode}% http://ctan.org/pkg/algorithmicx

\usepackage{subcaption}

\def\BibTeX{{\rm B\kern-.05em{\sc i\kern-.025em b}\kern-.08em
T\kern-.1667em\lower.7ex\hbox{E}\kern-.125emX}}

\newcommand{\bQ}{{\mathbf Q}}

\newcommand{\cA}{{\mathcal A}}
\newcommand{\cC}{{\mathcal C}}

\newcommand{\cR}{{\mathcal R}}
\newcommand{\cP}{{\mathcal P}}
\newcommand{\cS}{{\mathcal S}}

\newcommand{\dC}{{\mathbb C}}
\newcommand{\dE}{{\mathbb E}}
\newcommand{\dR}{{\mathbb R}}
\newcommand{\mH}{\mbox{$\bf H $}}

\newcommand{\mR}{\mbox{$\bf R$}}
\newcommand{\vb}{\mbox{$\bf b $}}
\newcommand{\vn}{\mbox{$\bf n $}}

\newcommand{\vr}{\mbox{$\bf r $}}
\newcommand{\vw}{\mbox{$\bf w $}}
\newcommand{\vy}{\mbox{$\bf y $}}

\newcommand{\ber}{\mbox{$\texttt{BER}$}}
\newcommand{\llr}{\mbox{$\texttt{LLR}$}}
\newcommand{\ex}{\textrm{exp}}
\newcommand{\opt}{\textrm{opt}}

\newcommand{\be}{\begin{equation}}
\newcommand{\ee}{\end{equation}}

\DeclareMathOperator*{\argmax}{arg\,max} % thin space, limits underneath in displays
\DeclareMathOperator*{\argmin}{arg\,min} % thin space, limits underneath in displays

\begin{document}
\title{Deep reinforcement learning approach to MIMO precoding problem: Optimality and Robustness}
\author{
\IEEEauthorblockN{Heunchul Lee, Maksym Girnyk and Jaeseong Jeong}
\IEEEauthorblockA{\\ \textit{Ericsson Research}, \textit{Ericsson AB}, Stockholm, Sweden}
}
\maketitle
% ------------------------------------------------------------------------------ abstract ------------------------------------------------------------------------------%

\begin{abstract}
In this paper, we propose a deep reinforcement learning (RL)-based precoding framework that can be used to learn an optimal precoding policy for complex multiple-input multiple-output (MIMO) precoding problems.
We model the precoding problem for a single-user MIMO system as an RL problem in which a learning agent sequentially selects the precoders to serve the environment of MIMO system based on contextual information about the environment conditions, while simultaneously adapting the precoder selection policy based on the reward feedback from the environment to maximize a numerical reward signal. 
We develop the RL agent with two canonical deep RL (DRL) algorithms, namely deep Q-network (DQN) and deep deterministic policy gradient (DDPG).
To demonstrate the optimality of the proposed DRL-based precoding framework, 
we explicitly consider a simple MIMO environment for which the optimal solution can be obtained analytically and show that DQN- and DDPG-based agents can learn the near-optimal policy to map the environment state of MIMO system to a precoder that maximizes the reward function, respectively, in the codebook-based and non-codebook based MIMO precoding systems. 
Furthermore, to investigate the robustness of DRL-based precoding framework, we examine the performance of the two DRL algorithms in a complex MIMO environment, for which the optimal solution is not known. The numerical results confirm the effectiveness of the DRL-based precoding framework and show that the proposed DRL-based framework can outperform the conventional approximation algorithm in the complex MIMO environment. 
\end{abstract}


\footnotetext[1]{Parts of this work have been presented at IEEE ICC 2020\cite{lee:20}.  In addition, this work has been submitted to the IEEE for possible publication. Copyright may be transferred without notice, after which this version may no longer be accessible.}
%\footnotetext[1]{Parts of this work have been presented at the IEEE International Conference on Communications, June 2020\cite{lee:20}.}

% ------------------------------------------------------------------------------ introduction section ------------------------------------------------------------------------------%
\section{Introduction}
% ------------------------------------------------ 5G/6G: massive and heterogeneous
% Motivation
\subsection{Recent trends in wireless industry}
The area of cellular communications is undergoing a revolutionary transformation, penetrating ever wider segments of society and industry. For instance, next-generation wireless communications are envisioned as a prime enabler for the fourth industrial revolution, Industry 4.0. Leveraging ubiquitous wireless connectivity, industrial automation and factory flexibility become possible at an unprecedented scale. To support such emerging services, we expect that future wireless networks will accommodate more stringent requirements on data rate, reliability, latency, availability and energy efficiency, bringing significant challenges for future wireless networks. Therefore, such future network architectures are expected to be too complex to be analyzed and optimized by conventional theoretical approaches. In order to match the stringent requirements, future networks should ensure self-organizing and self-optimizing capabilities by embracing \emph{artificial intelligence} (AI) as a new enabler, which will be a great step towards the future sixth-generation (6G) radio access technology \cite{Wikstroem:20}.

In fact, the 3rd generation partnership project (3GPP) standards group has been developing an artificial intelligence function, called network data analytics function (NWDAF), that provides network data analytics to network functions in the 5G core network, including software-defined networking (SDN) and network function virtualization (NFV) \cite{3gpp:23.501}. Apart from the AI applications into the core network, AI is being studied and applied to improve performance and reliability of the cellular radio access network (RAN). To date, machine learning (ML) has been widely proposed for upper-layer designs such as cell-association, scheduling and spectrum management \cite{Zhao:19} \cite{Yan:19} \cite{Challita:18}\cite{Chen:17}.
Traditional physical-layer design methods have an inherent limitation of relying on mathematical modelling of communication channels and systems, which becomes an issue in the future networks with challenging environments. This leads to the need of a new physical-layer paradigm based on ML algorithms that can learn and adapt the transmission strategies according to the actual observed environments. In \cite{OShea:17}, an autoencoder is trained in supervised learning (SL) for optimizing an end-to-end system performance.
Supervised and unsupervised learning approaches have been applied to hybrid precoding problems in multiple-input multiple-output (MIMO) systems in \cite{Huang:19}. 
In this paper, we present a reinforcement learning (RL) approach to the precoding problem in a practical MIMO-orthogonal frequency-division multiplexing (OFDM) system.

% ------------------------------------------------ Deep RL
%% Figure Texa
\begin{figure}%[htbp]
\centerline{\includegraphics[trim=0cm 0.5cm 0cm 0.5cm,clip,width=0.7\linewidth]{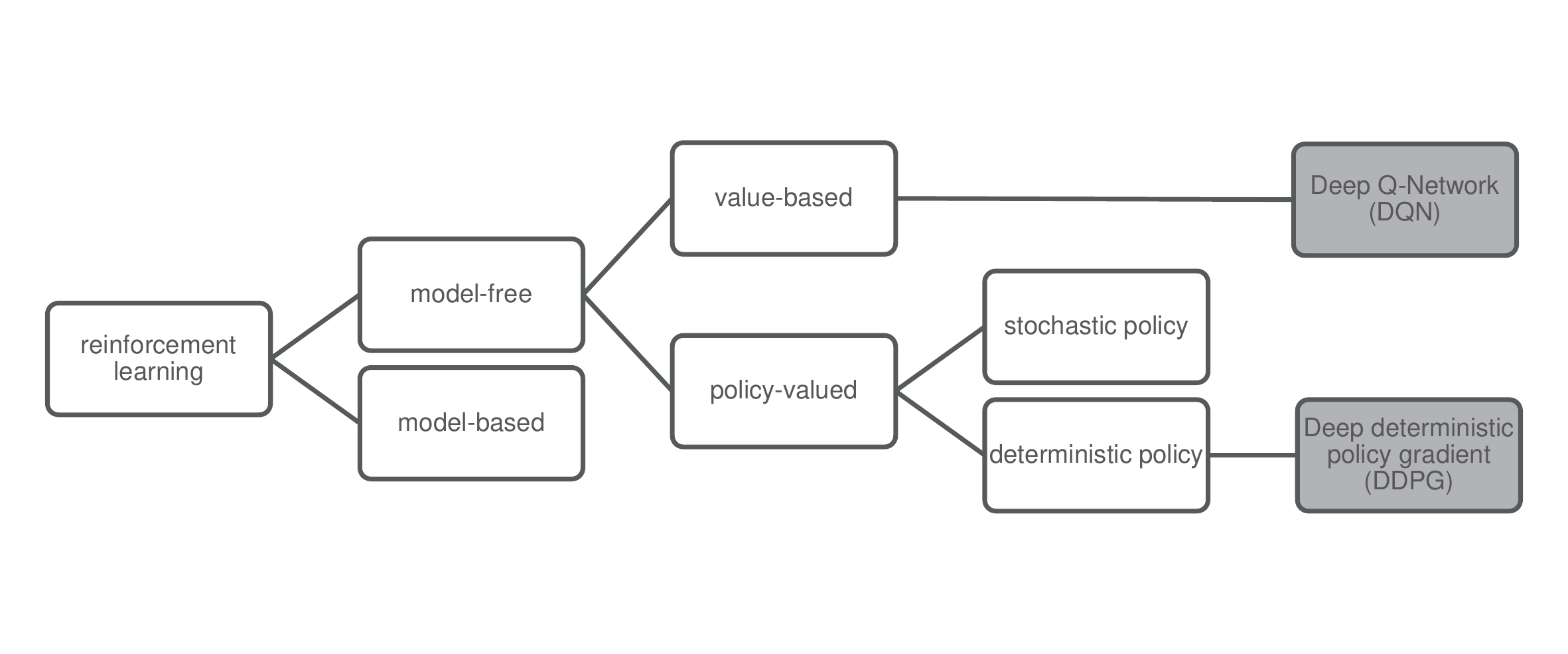}}
\caption{Model-free valued-based and policy-based reinforcement learning algorithms.}
\label{fig:texa}
\end{figure}
\subsection{Recent advances in machine learning}
%ML and DL+RL
ML, as a sub-field of AI, is playing an increasingly important role in many applications, ranging from small devices, such as smartphones and wearables, to more sophisticated intelligent systems, such as self-driving cars, robots and drones. RL is a set of ML techniques that allow an agent to learn an optimal action policy through trial-and-error interactions with a dynamic environment that returns the maximum reward \cite{Sutton:17}. These ML techniques are particularly relevant to applications where mathematical modelling and efficient solutions are not available. 
Figure \ref{fig:texa} shows the main subcategories of RL algorithms considered in this paper. RL algorithms can be classified into model-based and {\it model-free} methods, and the model-free methods can be further divided into {\it value-based} and {\it policy-based}. Model-based RL algorithms have access to a model of the environment or learn it. The environment model allows the RL agent to plan a policy by estimating the next state transitions and corresponding rewards. AlphaZero~\cite{Silver:2018} is an example of this category and uses a Monte-Carlo tree search model to predict the next series of state transitions, which in turn provides the value of each action before its execution. In contrast, model-free RL algorithms require no knowledge of state transitions and reward dynamics. These RL algorithms directly learn a value function or optimal policy from interactions with complex real-world environments, without explicitly learning the underlying model of the environment. Thanks to their easy implementation, model-free approaches have been widely used for a variety of applications. 
Motivated by recent advances in deep-learning (DL) \cite{Goodfellow:16}, deep reinforcement learning (DRL) combines the merits of DL with a RL learning model to achieve fully automated learning of optimal action policies. Deep Q-network (DQN) \cite{Mnih:13} and deep deterministic policy-gradient (DDPG) \cite{Lillicrap:16} are two leading model-free DRL algorithms to deal with discrete and continuous action space, respectively.

% ------------------------------------------------ Problem statement
\subsection{Problem statement and goal}
% DRL to MIMO
In this paper, we consider model-free RL algorithms for solving complex optimization problems in the physical-layer of wireless RANs. 
In particular, we investigate a DRL-based precoding framework for MIMO that remains as a key technology in future wireless networks. 
The use of multiple antennas at both transmitter and receiver in wireless communication links provides a means of achieving higher data rate and lower bit error rate (BER). 

The full potential of MIMO systems can be realized by utilizing channel state information (CSI) in the precoding design at the transmitter. The 3GPP 4G long-term evolution (LTE) and 5G new radio (NR) communication systems support two precoding modes, namely {\it codebook-based} and {\it non-codebook-based} precoding. In the codebook-based mode, the codebook consists of a number of predefined precoders, and is known at the transmitter and the receiver. The CSI is only known at the receiver, which chooses the index of the best precoder in the codebook and reports it to the transmitter. However, this precoding mode suffers from performance loss due to limited action space by the pre-defined codebook.
Meanwhile, the non-codebook based precoding mode can operate in a continuous action space in terms of precoders, trying to match the precoder to the actual channel realization. In this mode, the CSI is usually acquired from the channel reciprocity, and the precoder is computed based on the acquired CSI at the transmitter, while the receiver is not aware of the transmitter's choice of precoder.

% Problem statement & gap in knowledge
OFDM is another key technique adopted in modern communication systems. The multicarrier technique divides the total available bandwidth into a number of equally spaced subcarriers. The property of OFDM modulation 
turns a frequency-selective MIMO channel into a set of frequency-flat frequency-time resource elements (REs). An optimal precoding scheme would involve designing the best possible channel-dependent precoder on a per-RE basis. 
However, this approach is not practical due to issues with channel estimation and hardware implementation that arise on such a fine granularity. To achieve a tradeoff between performance and complexity in the design of practical MIMO-OFDM systems, a set of contiguous subcarriers are grouped into a so-called subband, and all the REs in each subband apply the same precoder, 
which is usually called subband precoding. A practical subband-precoding solution is obtained based on a spatial channel covariance matrix averaged over the pilot signals in a given subband. Unfortunately, this solution is sub-optimal, and furthermore no truly optimal solution has been known for this setting to date.

% Proposed DRL
To address this gap, we propose a DRL-based precoding framework that can be used to learn an optimal precoding policy for the subband precoding problem. We develop the RL agent with the two DRL algorithms: DQN and DDPG. DQN is a value-based RL algorithm that can only work in finite discrete action space, while DDPG is a policy-based RL algorithm that operates well in continuous action space. The value-based DQN solves RL problems with a finite set of actions by leveraging deep neural networks to estimate the action values for these actions. Thus, DQN is a natural fit to codebook-based precoding. In contrast to the DQN that derives an optimal policy indirectly through learning the optimal action-value function, the policy-based DDPG directly learns the policy in a continuous action space by updating the neural network parameters of a deterministic policy, following the deterministic policy gradient (DPG) algorithm~\cite{Silver:14}. Therefore, DDPG can be used to learn an optimal precoding policy in non-codebook based precoding mode.
To this end, we model the precoding problem for a single-user MIMO system as a RL problem in which a learning agent sequentially selects the precoders to serve the environment of MIMO system based on contextual information about the environment conditions. Meanwhile, the agent improves the precoder selection policy by adapting it based on the reward feedback from the environment to maximize a numerical reward signal. 

% ------------------------------------------------Main contribution
\subsection{Main contributions}
In order to demonstrate the optimality of the DRL-based precoding framework, 
we first consider a MIMO environment that consists of a MIMO-OFDM system with wideband precoding application and a flat-fading MIMO channel model, for which the optimal solution can be obtained analytically and show that DQN and DDPG-based agents can learn the near-optimal \footnote{Here, we refer to a policy that exhibits a limited gap in performance to the truly optimal policy.} policy for the precoder selection problem, respectively, in the codebook-based and non-codebook precoding modes, that maps the environment state (or CSI) of the MIMO system to an optimal action (or precoder) that maximizes a reward function. 
Furthermore, we investigate the robustness of the DRL-based precoding framework in learning a solution in very complex MIMO environments for which the optimal solution is not known. 
For this purpose, we evaluate the performance of the two DRL methods in a MIMO-OFDM system with subband precoding application and a frequency-selective MIMO channel model. The numerical results verify the effectiveness of the two DRL methods and show that the proposed precoding framework can outperform the conventional approximation algorithm in the complex MIMO environment. 

% ---------------------------------------- outline ------------------------------------------------------------ % 
The organization of the paper is as follows: 
Section \ref{sec:2} presents a system model of Rayleigh-fading MIMO-OFDM system and describes the precoding problems in codebook-based and non-codebook-based precoding modes. 
In Section \ref{sec:3}, we present a brief overview of fundamental concepts of deep RL algorithms
and describe the details of the two leading deep RL algorithms: DQN and DDPG.
In Section \ref{sec:4}, we present a DRL-based precoding framework for MIMO precoding problems and examine the performance of the proposed framework in the two different environments and show its optimality and robustness. 
Finally, conclusions are made in Section \ref{sec:5}.

% ------------------------------------------------------------------------------ system model and problem ------------------------------------------------------------------------------%
%% Figure MIMO
\begin{figure}%[htbp]
\centerline{\includegraphics[trim=0cm -2cm 0cm 0cm,clip,width=1.0\linewidth]{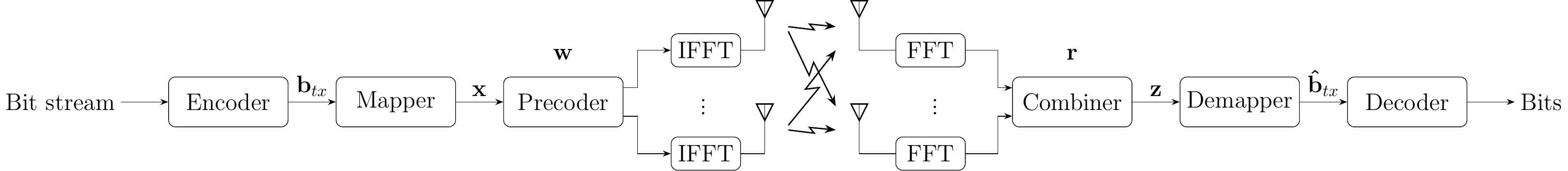}}
\caption{Schematic block diagram of MIMO-OFDM system, where a precoding vector $\vw \in \dC^{n_{tx}}$ and a combining vector $\vr \in \dC^{n_{rx}}$ are applied, respectively, on per-subband at the transmitter and on per-RE basis at the receiver.}
\label{fig:mimo}
\end{figure}
\section{System model and precoding problem} \label{sec:2}
This paper considers a wireless environment of a Rayleigh-fading MIMO-OFDM communication system with $n_{tx}$ transmit and $n_{rx}$ receive antennas. The system is assumed to exploit bit-interleaved coded modulation (BICM) that has been utilized in a wide range of wireless communication systems including the IEEE local-area network (LAN) and 3GPP LTE Systems \cite{lee:04}. A simplified block diagram of the BICM MIMO-OFDM system is presented in Fig. \ref{fig:mimo}, where a precoding vector $\vw \in \dC^{n_{tx}}$ and a combining vector $\vr \in \dC^{n_{rx}}$ are applied, respectively, per-subband at the transmitter and on a per-RE basis at the receiver for exploiting the spatial diversity available in MIMO channels. 
At the transmitter, one transport bit stream is encoded to a bit block $\vb_{tx}$ which is then symbol-mapped to modem symbols $x$. Typical modem constellations used are $M$-QAM, consisting of a set of $M$ constellation points, the set being denoted by $\cC$. Then, the data symbols $x$ are precoded by the precoding vector $\vw$ to form $n_{tx}$ data substreams. Finally, the substreams are transmitted via the available multiple transmit antennas.

%% Figure PRB
\begin{figure}%[htbp]
\centerline{\includegraphics[trim=0cm 0cm 0cm 0cm,clip,width=0.8\linewidth]{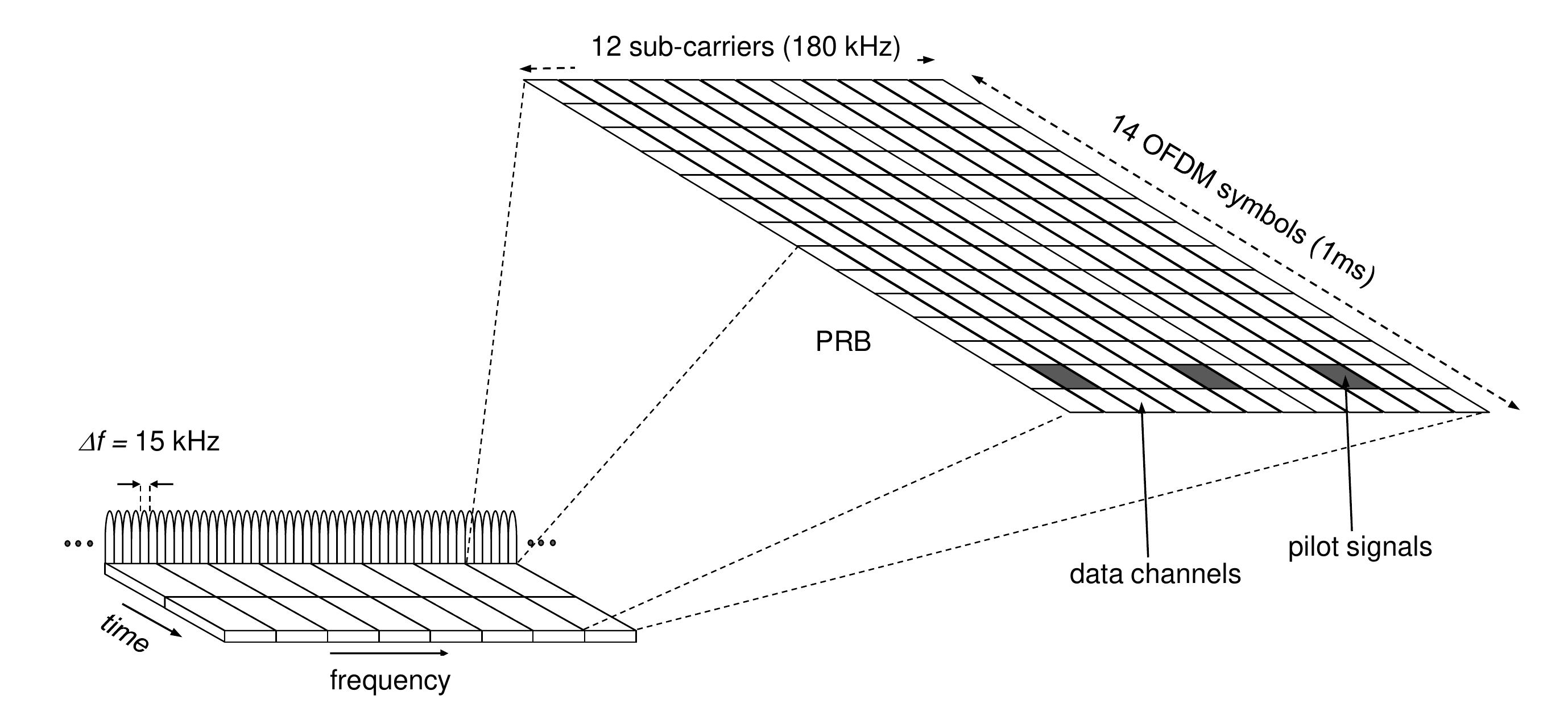}}
\caption{An illustrative example of time-frequency resource grid of MIMO-OFDM system with resource parameters specified in the 3GPP standards \cite{3gpp:36.211}.}
\label{fig:prb}
\end{figure}

In NR systems, the smallest available physical resource unit is a \emph{resource element} (RE). One RE consists of one OFDM subcarrier in frequency domain and one OFDM symbol in time domain. A group of REs forms a \emph{physical resource block} (PRB) which is the basic resource allocation unit. Figure \ref{fig:prb} illustrates an example of PRB in time-frequency resource grid of MIMO-OFDM system with resource parameters specified in the 3GPP standards \cite{3gpp:36.211}. Assuming a subcarrier spacing of $\Delta f$= 15kHz, each PRB is formed by 12 consecutive OFDM subcarriers, 180kHz wide in frequency, and 14 OFDM symbols, 1ms long in time. This transmission duration of 1ms corresponds to one transmission time interval (TTI) specified in the LTE standards. 
As can be seen in Fig. \ref{fig:prb}, one set of REs, called data channels, is used to carry information originating from the higher layers, while the other set, called pilot signals, conveys reference signals for channel estimation.

Focusing on the data transmission in the forward link from the transmitter to the receiver, let $x_i$ denote the complex data symbol at the $i$-th RE on the forward link data channels and let $\vy_i $ be the corresponding $n_{rx}$-dimensional complex received signal vector, which can be written as 
\be
\vy_i =\mH_i \vw x_i +\vn_i, 
\ee 
where $\mH_i \in \dC^{n_{rx}\times n_{tx}}$ represents the MIMO channel matrix between transmit and receive antennas, and $\vn_i \in \dC^{n_{rx}}$ is an additive white Gaussian noise (AWGN) vector whose elements are independent identically distributed (i.i.d.) complex-valued Gaussians with zero mean and variance $\sigma_n^2$. Under the Rayleigh-fading channel, the channel matrix $\mH_j$ is represented as
\be
\mH_j= \begin{pmatrix}
h_{j,1,1} & h_{j,1,2} & \dots & h_{j,1,n_{tx}} \\
h_{j,2,1} & h_{j,2,2} & \dots & h_{j,2,n_{tx}} \\
\vdots & \vdots & \ddots & \vdots \\
h_{j,n_{rx},1} & h_{j,n_{rx},2} & \dots & h_{j,n_{rx},n_{tx}}
\end{pmatrix},
\ee
where $h_{j,m,n}$ represents the channel coefficient from the transmit antenna $n$ to the receive antenna $m$ at the RE $j$, and the channel elements are obtained from an i.i.d. complex Gaussian distribution with zero mean and unit variance.

Without loss of generality, we assume that the data symbol $x_i$ and the precoding vector $\vw$ are normalized as follows: $\dE\left[ |x_i|^2 \right]=1$ and $\|\vw\|=1$, where $\dE \left[ \cdot \right]$
denotes the expectation with respect to the distribution of the underlying random variable, and $|\cdot|$ and $\|\cdot\|$ denote, respectively, the absolute value of a complex number and the 2-norm of a vector.
Under these assumptions, the signal-to-noise ratio (SNR) $\rho$ is given by $\rho=1/\sigma_n^2$. 

At the receiver, the transmitted data symbol $x_i$ can be recovered by combining the received symbols $\vy_i$
by the unit-norm vector $\vr_i$ (i.e., $\|\vr\|=1$), which yields the estimated complex symbol
\be \label{eq:chgain}
z_i =\vr_i^{\dag}\vy_i =\vr_i^{\dag}\mH_i \vw x_i + \vr_i^{\dag}\vn_i.
\ee

Note that $\vr_i^{\dag}\mH_i \vw$ in \eqref{eq:chgain} corresponds to the
effective channel gain. We assume a maximal ratio combiner (MRC) is used at the receiver, given by
\be 
\vr_i=\frac{\mH_i~\!\vw}{\|\mH_i~\!\vw\|},
\ee
which is optimal in the sense of output SNR maximization when the noise is white.

\subsection{Precoding problems in codebook-based and non-codebook based MIMO systems}
In this section, we describe the codebook-based and non-codebook based MIMO precoding systems.
In particular, we address the MIMO precoding design on a per-subband basis, which is representative of MIMO precoding applications in real-world deployments. 
We consider a subband-based precoding, where each subband is formed by a certain number of consecutive PRBs and the precoder $\vw$ is the same for all the data channels within a subband.
We denote the set of data REs and pilot REs in a given subband, respectively, by $\Phi_{d}$ and $\Phi_{p}$ in the forward link and by $\Psi_{d}$ and $\Psi_{p}$ in the reverse link. 

As BICM is adopted in 3GPP LTE and NR Systems, we investigate the precoding problem for optimizing the BER performance.
We calculate the uncoded BER performance by comparing the transmit bit block $\vb_{tx}$ and the receive bit block $\hat{\vb}_{tx}$ as they represent the value of precoder $\vw$ over the MIMO channel without the help of channel coding.
The data bit block $\vb_{tx}$ can be recovered from \eqref{eq:chgain}. 
Let $b_i^m$ be the bit that is mapped into the $m$-th bit position ($m=1,2,\ldots,\log_2M$) of the constellation symbol $x_i$.
Then, the log-likelihood ratio (LLR) value of $b_i^m$ can be defined as \cite{lee:04}
\be \label{eq:llr}
\llr(b_i^m) = \log \frac{\sum_{x\in \cC_0^m } \exp{\left(-\frac{\|z_i - \vr_i^{\dag}\mH_i\vw x \|}{\sigma_n^2}\right)}}{\sum_{x\in \cC_1^m } \exp{\left(-\frac{\|z_i - \vr_i^{\dag}\mH_i\vw x\|}{\sigma_n^2}\right)}},
\ee
where the set $\cC_d^m$, $d=0$ or $1$, is the set of all symbols $x_i$ in the constellation set $\cC$ with $b_i^m=d$.

% reciprocity
Under channel reciprocity, we assume that the forward CSI is available for the transmitter to compute the precoding vector $\vw$.
For the setup described above, the transmitter can obtain the forward CSI by estimating the reverse CSI via the reverse pilot channels on the RE set $\Psi_{p}$. 
Therefore, the environment state of the MIMO-OFDM system is defined by a set of MIMO channel matrices $\mH_j$ on the forward-link REs $j$ that correspond to the reverse-link set $\Psi_{p}$. More specifically, the forward channel matrix $\mH_j$ on the RE $j \in \Psi_{p}$ is given by the transpose of the MIMO channel matrix estimated on the corresponding reverse pilot signal $j$. 
The forward-link precoder $\vw$ is derived based on the MIMO matrices $\left\{ \mH_j \right\}_ {j \in \Psi_{p}}$ and the same precoder $\vw$ is applied for forward-link data transmission on all the data channels $i\in \Phi_{d}$.

The conditional BER performance of action $\vw$ from a given environment state $\{\mH_j\}_{j\in \Psi_{p}}$ can be defined as
\be \label{eq:ber}
\ber\left(\vw |\{\mH_j\}_{j\in \Psi_{p}}\right) \triangleq \frac{1}{\log_2M} \sum_{m=1}^{\log_2M}\dE_{\{x_i,\vn_i\}_{i\in \Phi_{d}}}\left[b_i^m \ne \hat{b}_i^m |\{\mH_j\}_{j\in \Psi_{p}}\right],
\ee
where $\hat{b}_i^m$ denotes the hard decision bit of ${b}_i^m$ at the receiver, which is given by $\hat{b}_i^m=0$ if $\llr(b_i^m) > 0$, and by $\hat{b}_i^m=1$ otherwise. 

Finally, the optimal precoder can be obtained by the following BER minimization problem 
\be \label{eq:bermin}
\vw^{\opt}=\argmin_{\vw \in \cA} \; \ber\left(\vw|\{\mH_j\}_{j\in \Psi_{p}}\right).
\ee
Here the action space $\cA$ is given by a discrete action space $\cA_d$ in codebook-based MIMO precoding systems, that consists of all precoders in a given codebook. Meanwhile, it is given by a continuous action space $\cA_c$ in non-codebook based MIMO precoding systems, spanned by all possible precoders under power constraint.

Unfortunately, the minimization problem in \eqref{eq:bermin} does not admit a computationally efficient solution. 
In the following sections, we consider a DRL-based approach as an alternative. That is, instead of computing an optimal precoder $\vw^{\opt}$ we are learning it through interactions with the environment of the MIMO-OFDM system.

% ------------------------------------------------------------------ Deep RL approach -------------------------------------------------------------------%

\section{Deep reinforcement learning} \label{sec:3}
In this section, we first present a brief overview of the fundamental concepts in DRL, including action value, Q-learning, and function approximation, and describe details of 
DQN and DDPG.

% RL and MDP
%% Figure RL MDP
\begin{figure}%[htbp]
\centerline{\includegraphics[trim=0cm -1cm 0cm -1cm,clip,width=0.7\linewidth]{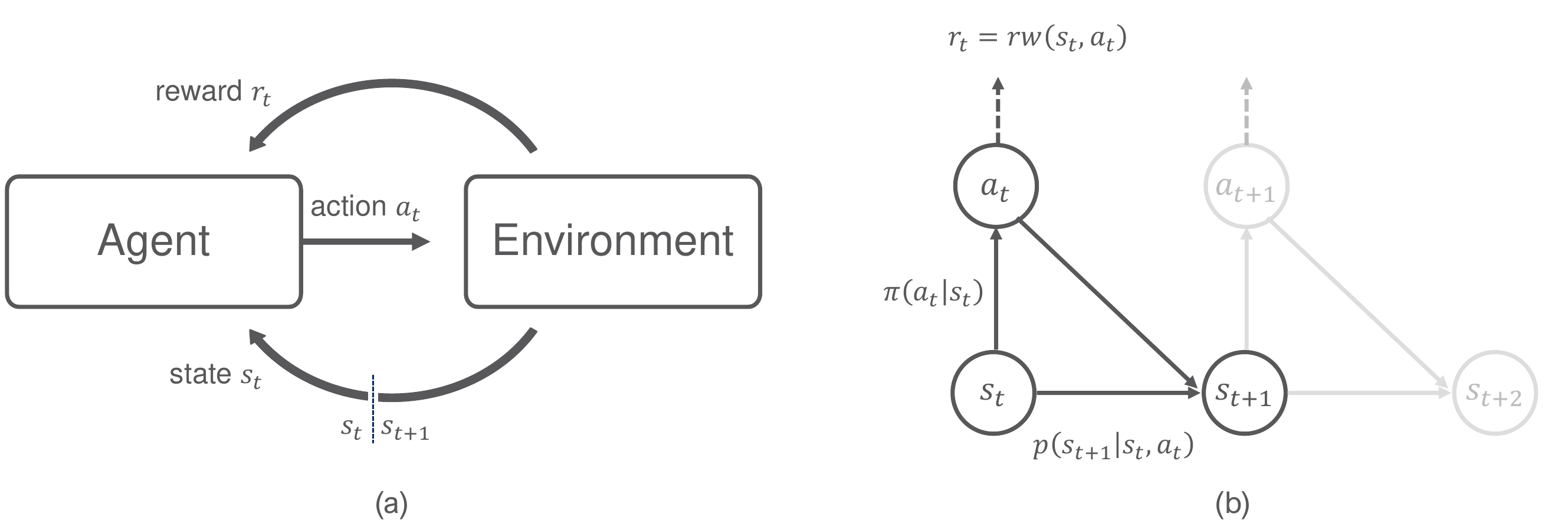}}
\caption{(a) Reinforcement learning through interactions between agent and environment and (b) The interaction between agent and environment can be modeled as Markov-decision process.}
\label{fig:rlmdp}
\end{figure}
RL is a set of ML techniques that allows an agent to learn the optimal action policy that returns the maximum reward through trial-and-error interactions with a challenging dynamic environment \cite{Sutton:17}.
Figure \ref{fig:rlmdp} illustrates reinforcement learning through interactions between the agent and the environment. Most RL problems can be formalized by modelling the interaction between the agent and the environment as a \emph{Markov decision process} (MDP). An MDP consists of a set of environment states $\cS$, a set of available actions $\cA$, a stochastic reward function $\cR$ and a state transition function $\cP: \cS \times \cA \rightarrow \cS$ from one state to another given an action taken. 
A policy is a mapping function from state to action in the MDP that specifies action $a$ that is taken in state $s$. In general, the policy is stochastic and denoted by the conditional distribution $\pi\left(a|s\right)$ while a deterministic policy is specifically denoted by $a=\mu(s)$ to emphasize a deterministic function.

The random variables, $s_t \in \cS$, $a_t \in \cA $, and $r_t \in \cR$ denote the state, action and reward values at time step $t$, where the reward $r_t$ is a function of a state–action pair $\left(s_t,a_t\right)$, denoted by $rw\left(s_t,a_t\right)$, in MDP. 
At each time step $t$, an agent observes a state $s_t$ of environment conditions and chooses an action $a_t$ to serve the environment. After each time step, the agent gets an immediate reward $r_t$ and next state $s_{t+1}\in\cS$ in return for the action taken. 
Then, each experience transition at $t$ can be represented by the tuple
\be \label{eq:ex}
e_t=\left[s_t,a_t,r_t,s_{t+1}\right].
\ee
The agent aims to maximize the future cumulative return $R_t$ from time step $t$ onwards defined as 
\be
R_t=\sum_{k=t}^{\infty}\gamma^{k-t}r_t,
\ee
where $\gamma\in\left[0,1\right]$ denotes a discounting factor to the future rewards.

The state-action value, called Q-value, of a state-action pair $(s, a)$ is defined as the expected return achievable by an action $a$ in a state $s$ by following policy $\pi$
\be 
\bQ^{\pi} (s, a )=\dE \left[R_0|s_0=s, a_0=a,\pi \right],
\ee
where $s_0$ denotes the initial state and
the expectation is taken over all the possible state-action transitions given by policy $\pi$.

The agent's goal can be achieved by finding the optimal policy $\pi^*$ that returns the maximum expected cumulative reward at each state 
\be
\pi^*=\argmax_{\pi } \bQ^{\pi} (s, a).
\label{eq:objRL}
\ee

This goal can be achieved by different types of RL algorithms shown in Fig.~\ref{fig:texa}.
In what follows, we describe the two model-free algorithms: value-based algorithm and policy-based algorithm
that provides a basic framework for learning an optimal precoding policy in a codebook-based and non-codebook based precoding mode,
respectively.

% ------------------------------------------- %
\subsection{Value-based RL: DQN}
We first describe the DQN algorithm using an action selection known as $\epsilon$-greedy. For the completeness of presentation, this section presents a summary of deep Q-learning algorithm from \cite{Sutton:17}, \cite{Mnih:13},\cite{Mnih:15}, and \cite{Kaelbli:96}. 

In value-based RL algorithms, the RL problem is solved by estimating the optimal value of each action when taking that action for a given state.
The agent learns the value function through trial-and-error interactions with the environment until it converges to the {\it optimal} Q-function $\bQ^{*} (s, a )$ corresponding to the optimal policy. 
From this definition, a simple optimal strategy is obtained to take the action $a^*$, called {\it greedy action}, with the highest action value in given state $s$ as follows: 
\be \label{eq:action}
a^* =\argmax_{a} \bQ^{*} (s, a ).
\ee

The basic idea of the off-policy Q-learning algorithm is to approximate the optimal value function $\bQ^{*} (s, a )$ by a tabular representation $\bQ_{T}$ or a function approximator $\bQ _{m}$ with a specific model $m$.
The tabular method $\bQ_{T}$ is the simplest form of Q-learning that can generate a Q-table $T$ with all possible state-action pairs when the state and action spaces are small enough for the tabular representation. 
However, the problem with a table-based Q-learning approach is that the training complexity and the memory requirements become too large in the typical optimization tasks for wireless communications since all the possible state-action pairs $(s, a)$ should be visited to update the Q-values, denoted by $\bQ_{T}(s, a)$.

%% Figure DQN
\begin{figure}%[htbp]
\centerline{\includegraphics[trim=0cm -2cm 0cm 0cm,clip,width=0.5\linewidth]{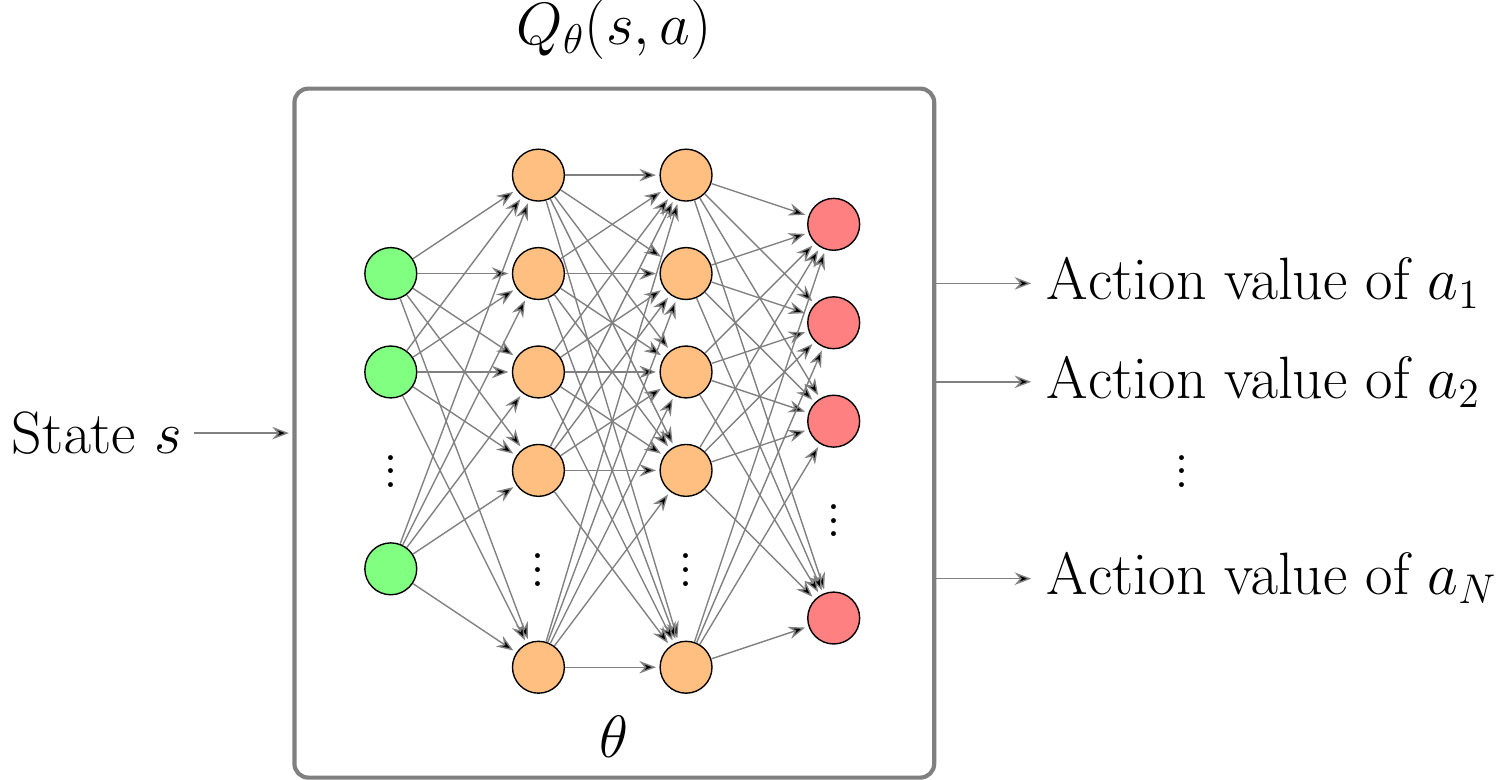}}
\caption{A schematic illustration of the value-based DQN with parameters $\theta$ that can be used for solving the codebook-based precoding problem in the proposed precoding framework, assuming a discrete action space given by $\cA_d=\left\{a_1,a_2,\cdots,a_N\right\}$.}
\label{fig:dqn}
\end{figure}
% DQN
Instead, the DQN algorithm utilizes a deep neural network with parameters $\theta$ as a generalizing function approximator in the Q-learning \cite{Kaelbli:96}. 
Figure \ref{fig:dqn} illustrates a DQN function approximator , denoted by $\bQ _{\theta}$, that takes state $s$ as an input and produces a separate output for each action $a \in\cA_d$. 
The use of neural network in Q-learning has the benefit of generalization over the continuous state spaces that the agent can perform well in testing environments similar to the environments that it has seen before during learning \cite{Goodfellow:16}. This means that DQN can produce a good approximation over the entire state space by learning only with a limited subset of the state space. 
Therefore, the DQN algorithm can find the approximate value functions effectively even for much larger problems with multidimensional and continuous states, while suffering less from the curse of dimensionality compared to the tabular method. 

The optimal Q-function can be represented, by using the iterative \emph{Bellman optimality equation} \cite[Ch. 11]{Sutton:17} \cite{Bellman:57}, as
\be \label{eq:bell}
\bQ^{*} (s, a ) = \dE \left[r_t+\gamma \max_{\hat{a}\in\cA_d} \bQ^{*} (s_{t+1}, \hat{a} ) |s_t=s, a_t=a\right]. 
\ee
The Bellman equation provides a recursive definition for a temporal-difference (TD) based Q-learning algorithm for approximating the optimal Q-function by measuring the difference between the current Q-value estimate (referred to as online Q-value), and the new estimate (referred to as target Q-value).

%% Figure exploration
\begin{figure}%[ht]
\begin{subfigure}{.5\textwidth}
\centering
% include first image
\includegraphics[trim=0cm -0.5cm 0cm 0cm,clip,width=1.05\linewidth]{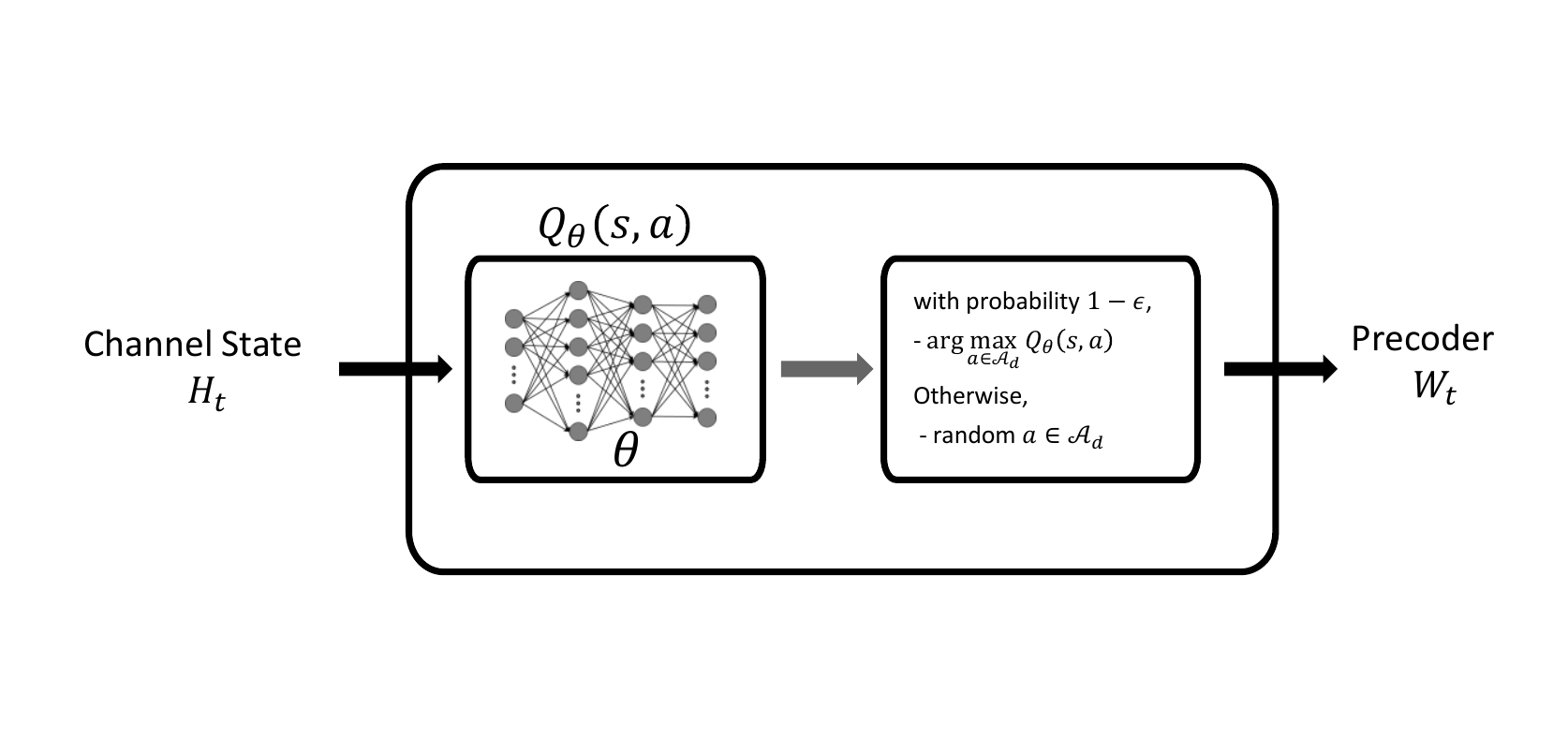}
\caption{DQN with an $\epsilon$-greedy strategy}
\label{fig:expl1}
\end{subfigure}
\begin{subfigure}{.5\textwidth}
\centering
% include second image
\includegraphics[trim=0cm -0.5cm 0cm 0cm,clip,width=1.05\linewidth]{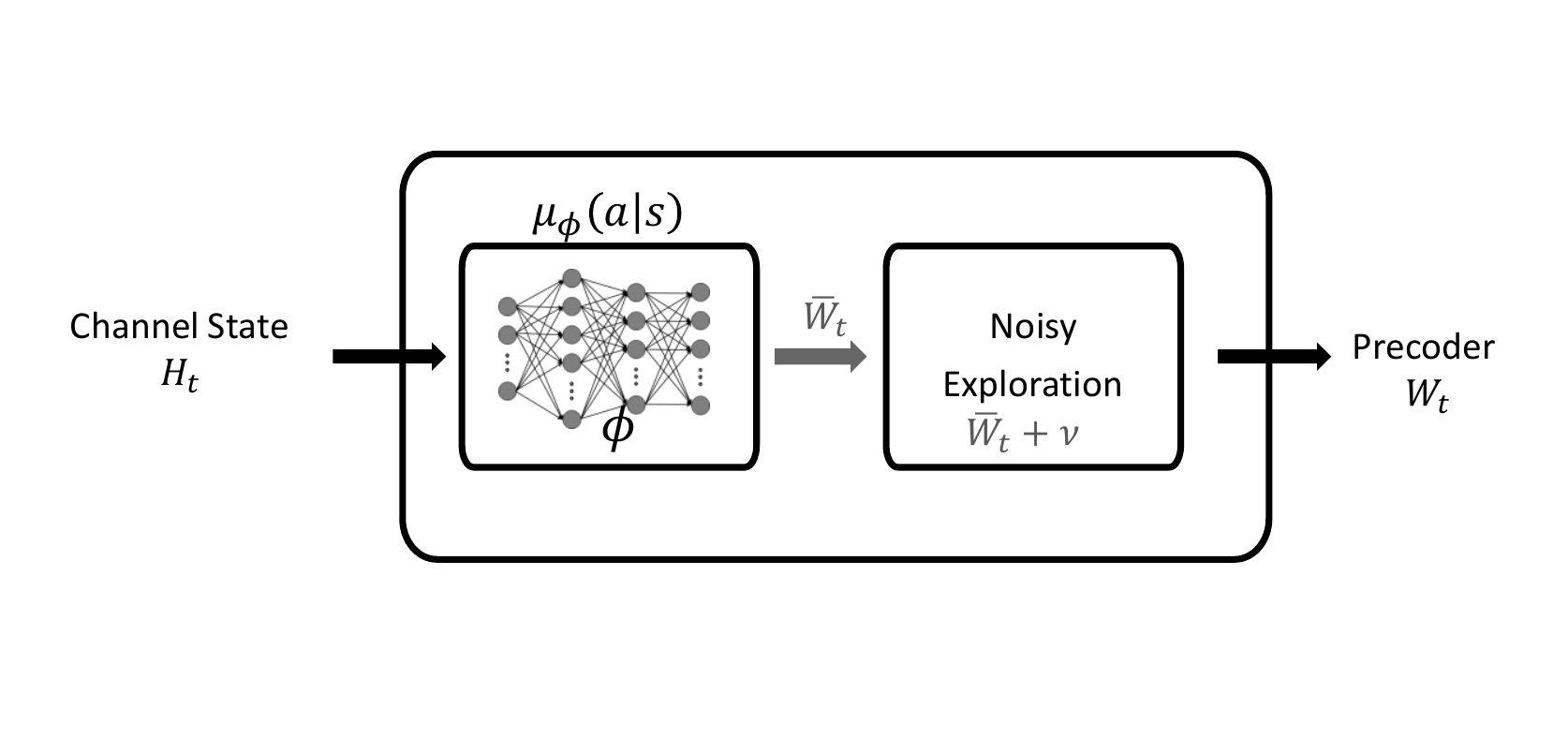}
\caption{DDPG with exploration noise}
\label{fig:expl2}
\end{subfigure}
\caption{Exploration strategies for behavior policies in the off-policy DQN and DDPG algorithms}
\label{fig:expl}
\end{figure}
Let $\bQ_{\theta} (s, a)$ denote a DQN function approximator $\bQ_{\theta}$ indexed by the state-action pairs $(s,a)$ to estimate the optimal action value $\bQ^{*} (s, a)$.
At the beginning of training, the network parameters $\theta$ are randomly initialized.
At each step $t$, the agent observes the state $s_t$ and selects an action $a_t$ from the pre-defined set $\cA_d$. 
In case of {\it off-policy} Q-learning, we learn the optimal {\it target} policy from experiences generated by a different policy, called {\it behavior} policy, that is used during exploration. Figure \ref{fig:expl1} illustrates the most popular exploration strategy for a behavior policy, called $\epsilon$-greedy strategy, which is represented as
\be \label{eq:egreedy}
a_t=\begin{cases}
\mbox{ with probability } \epsilon, \mbox{select a random action $a \in \cA_d$ } \\
\mbox{otherwise, take a greedy action by} \argmax_{a \in \cA_d} \bQ_{\theta} (s_t,a)
\end{cases}.
\ee
There are two issues related to the action selection: First, if an agent chooses a greedy action by the Q-value in (\ref{eq:egreedy}), the greedy action selection may result in local optimization problem. Second,
this online selection involves an exploration-exploitation dilemma, which is a fundamental trade-off between maximizing the expected immediate reward in the current step and achieving the greater total reward in the long run. 
Since exploration is costly due to the limited computational resources such as time and data,
the $\epsilon$-greedy method with decaying $\epsilon$ is applied to start with a high exploration rate and reduce it at each time step.
In other words, at the early stage of learning, the agent wants to explore more to learn the best policy based on trial and error, improving the overall Q-value estimates at the cost of the short-term sacrifices, and gradually exploits more to produce the maximum total reward. 

After each experience $\left[s_t,a_t,r_t\right]$, we can evaluate the online Q-value as 
\be \label{eq:old}
\bQ _{\theta}(s_t,a_t),
\ee
and calculate a new target Q-value, denoted by $Y_t^\theta$, according to the Bellman equation in (\ref{eq:bell}), as follows:
\be \label{eq:new}
Y_t^{\theta}=r_t + \gamma \max_{\hat{a}\in\cA_d} \bQ _{\theta}(s_{t+1}, \hat{a} ),
\ee 
where the value on the current step is expressed via the value of greedy action on the next step.

Accordingly, the loss function is defined as the squared error between the two values 
\be
L({\theta})=\frac{1}{2}\left| Y_t^{\theta} - \bQ _{\theta}(s_t,a_t) \right|^2. 
\ee

Then, DQN learns the optimal action value function $\bQ^{*} (s, a)$ by finding the optimal parameters $\theta$ through the loss minimization problem with respect to a loss function $L({\theta})$. A standard approach for the loss function optimization is the gradient descent algorithm. In practice, the true gradient decent is approximated by a procedure called stochastic gradient descent (SGD) to efficiently update the parameters.

The parameter update can be made by adjusting the parameters in the opposite direction of the gradient
\be
\theta \leftarrow \theta-\eta\triangledown_{\theta} L({\theta}),
\ee
where $\eta\in\left[0,1\right]$ is a learning rate and $ \triangledown_{\theta} L({\theta})$ denotes the gradient of the loss function with respect to the parameters.

The above update rule can be further expressed in the term of error between the current estimate in (\ref{eq:old}) and the new estimate in (\ref{eq:new}) as follows:
\be \label{eq:sgd}
\theta \leftarrow \theta+\eta\left(Y_t^{\theta} - \bQ _{\theta}(s_t,a_t ) \right) \triangledown_{\theta} \bQ _{\theta}(s,a ),
\ee
where $\triangledown_{\theta} \bQ _{\theta}(s,a)$ denotes the vector of partial derivatives with respect to the components of $\theta$.
By combining the gradient descent method with the backpropagation algorithm, we can update all the network parameters in the input, hidden and output layers \cite{LeCun:98}. 

In summary, the learned action-value function $\bQ _{\theta}$ directly approximates the optimal action-value function $\bQ^*$, and, as shown in (\ref{eq:action}), the optimal policy is to take the action that leads to the highest value in a given state. We also note that the off-policy DQN algorithm has an advantage of TD learning that allows the parameter update by adjusting from the current estimate to the more accurate estimate computed for each new experience $[s_t, a_t,r_t ]$. 

Throughout this paper, we have assumed to use the same Q-network to select and to evaluate an action in the computation of the target value in (\ref{eq:new}), which can lead to an overestimation problem in Q-learning due to a high correlation between the target value and the parameters being updated by (\ref{eq:sgd}). To handle this high correlation problem, we may utilize two separate Q-networks in the action selection and evaluation for improved Q-learning. Specifically, the double DQN (DDQN) makes the update with the target value given by two separate Q-networks, namely, DQN network $\bQ _{\theta_1}$ and target network $\bQ _{\theta_2}$, as follows \cite{Hasselt:16}:
\be
Y_t^{\theta_1,\theta_2}=r_t + \gamma \bQ _{\theta_2}(s_{t+1}, \arg \max_{\hat{a}} \bQ _{\theta_1}(s_{t+1}, \hat{a} ) ),
\ee
where the DQN network $\bQ _{\theta_1}$ is used in the action selection and the target network $\bQ _{\theta_2}$ is used for the action evaluation.

As seen in the equations \eqref{eq:action} and \eqref{eq:egreedy},
the greedy action policy in the training and execution phase becomes computationally intractable with continuous action space, and thus Q-learning algorithm can only work with discrete action sets $\cA_d$. In the next section, we review policy-based RL algorithms that can be applied to continuous action spaces.

% ------------------------------------------- %
\subsection{Policy-based RL: DDPG}
A review of basic policy gradient algorithms \cite{Sutton:17} is provided, followed by variants of the basic policy gradient algorithms that expand to deterministic policy gradient in \cite{Silver:14} and its deep learning version DDPG \cite{Lillicrap:16}. For readers with no prior background in the field, we will provide a brief essential background on DDPG (See \cite{Silver:14} and \cite{Lillicrap:16} for additional details).

The basic idea behind stochastic policy gradient algorithms is to move policy $\pi$ in the direction of the performance gradient. We consider a stochastic policy $\pi_{\phi}(a|s=s_t)$ with parameters $\phi$. In stochastic policy gradient algorithms, the action $a_t$ is sampled by the stochastic policy that determines the probability distribution $\pi_{\phi}(a|s=s_t)$ of all possible actions $a$ given a state $s_t$. We can optimize the policy $\pi_{\phi}$ by adjusting the policy parameters $\phi$ in the direction of the gradient of the expected reward
\be \label{eq:object}
J(\pi_{\phi})= \dE_{s,a\sim\pi_{\phi} }\left[rw\left(s_t,a_t\right)\right].
\ee
As opposed to gradient descent, the policy gradient algorithms work by updating policy parameters $\phi$ via a gradient ascent on policy as follows:
\be
\phi \leftarrow \phi+\eta\triangledown_{\phi} J(\pi_{\phi}),
\ee
where $\eta$ is a learning rate.

As can be seen in \eqref{eq:object}, we need to compute the gradient $\triangledown_{\phi} J(\pi_{\phi})$ over the action distribution and the state distribution both dependent on $\pi_{\phi}$, which is a challenging problem. The policy gradient theorem simplifies this computation by using the expectation of the product of the action value and gradient of the logarithm of the policy $\pi_{\phi}$ expressed as
\be \label{eq:policygradient}
\triangledown_{\phi} J(\pi_{\phi}) = \dE_{s,a}\left[ \triangledown_{\phi}\log{\pi_{\phi}\left(a|s\right)} \bQ^{\pi}(s,a)\right].
\ee

By the policy gradient theorem, the policy gradient does not depend on the gradient of the state distribution of $s$. Several policy gradient algorithms are proposed based on the policy gradient theorem. One challenge for these policy gradient algorithms is to find the true action-value function $\bQ^{\pi}(s,a)$ in \eqref{eq:policygradient}. 
A vanilla policy gradient method can be applied to estimate the true action-value function $\bQ^{\pi}(s,a)$ by using a sample return $R_{t:T}=\sum_{k=t}^{T}\gamma^{k-t}r_t$. The update rule based on the sample estimate $R_{t:T}$ is given by
\be
\phi \leftarrow \phi + \eta \triangledown_{\phi}\log{\pi_{\phi}\left(a_t|s_t\right)} R_{t:T}.
\label{eq:vanilla_pg}
\ee
Although this method provides an unbiased gradient estimate, the vanilla policy gradient method suffers from high variance
of gradient estimates due to the randomness of Monte-Carlo estimation $R_{t:T}$. 

To reduce the variance of the gradient estimator, an actor-critic method was introduced. It approximates $\bQ^{\pi}(s,a)$ in \eqref{eq:policygradient} by a trained critic function $\bQ_{\vartheta}$ with a parameter $\vartheta$, resulting in the gradient 
\be \label{eq:stochastic}
\triangledown_{\phi} J(\pi_{\phi}) = \dE_{s,a}\left[ \triangledown_{\phi}\log{\pi_{\phi}\left(a|s\right)} \bQ_{\vartheta}(s,a)\right],
\ee
where the substitution of the true function $\bQ^{\pi}$ by the function approximator $\bQ_{\vartheta}$ may introduce bias in the estimation as a penalty for the variance reduction.

We note that all the stochastic policy gradient algorithms described above in \eqref{eq:policygradient} to \eqref{eq:stochastic} are computationally expensive to implement for high dimensional, continuous action spaces because the gradient needs to be estimated over the entire state and action space. 

Deterministic policy gradient (DPG) algorithms extend the actor-critic idea from discrete to the continuous action space \cite{Silver:14}. 
As a deep variant of DPG, 
DDPG combines DPG with DQN in an actor-critic setting, that can operate in continuous action spaces \cite{Lillicrap:16}. DDPG consists of two deep neural networks: actor network $\mu_{\phi}$ parametrized by $\phi$ for approximating a deterministic policy and critic network $\bQ _{\vartheta}$ parametrized by $\vartheta$ for estimating the action-value function. DDPG algorithm uses the critic $Q_{\vartheta}(s,a)$ to estimate the optimal action-value function and updates the actor $\mu_{\phi}(s)$ in the direction of the gradient of $Q_{\vartheta}(s,\mu_{\phi}(s))$, which is given by
\be \label{eq:dpg1}
\triangledown_{\phi} J(\mu_{\phi}) = \dE_{s}\left[ \triangledown_{\phi} \bQ _{\vartheta}(s,a)|_{a=\mu_{\phi}(s)} \right].
\ee
Applying the chain rule to \eqref{eq:dpg1}, we have
\be
\triangledown_{\phi} J(\mu_{\phi}) = \dE_{s}\left[\triangledown_{\phi}\mu_{\phi}(s) \triangledown_{a} \bQ _{\vartheta}(s,a)|_{a=\mu_{\phi}(s)} \right],
\ee
where $\triangledown_{\phi}\mu_{\phi}(s)$ denotes the gradient of $\mu_{\phi}$ with respect to the parameters $\phi$ and $\triangledown_{a} \bQ _{\vartheta}(s,a)$ is the gradient of $\bQ _{\vartheta}$ with respect to the action $a$.

Finally, for each experience at time step $t$, we can update the parameters $\phi$ via a stochastic gradient ascent 
\be
\phi \leftarrow \phi+\eta \triangledown_{\phi}\mu_{\phi}(s)|_{s=s_t} \triangledown_{a} \bQ _{\vartheta}(s,a)|_{s=s_t,a=\mu_{\phi}(s_t)}.
\ee
Even though we use a deterministic policy $a=\mu_{\phi}(s)$ that always yields the same action for the same state, a stochastic policy is desirable for exploration in the learning phase. Fortunately, as the DDPG is an off-policy algorithm, we can treat the exploration problem independently in the algorithm. For exploration, DDPG uses a stochastic behavior policy to select actions, as will be described in Section \ref{sec:4}. 
In summary, the DDPG is a model-free off-policy actor-critic algorithm that learns the critic $\bQ _{\vartheta}$ directly using the experience samples generated by a stochastic behavior policy and also learns about the target policy $\mu_{\phi}$ directly from the value function $\bQ _{\vartheta}$.

% ------------------------------------------------------------------ DQN and DDPG: optimality and robustness -------------------------------------------------------------------%
\section{Deep RL-based precoding framework: Optimality and Robustness} \label{sec:4}
In this section, we present a DRL-based precoding framework for MIMO precoding problems described in Section \ref{sec:2}
and investigate the learning performance of the proposed approach in terms of BER performance, compared to conventional precoding solutions.
In the proposed precoding framework, the RL agent interacts with an environment of MIMO system and channel by observing channel states, choosing precoders, getting BER performance over time steps. Through such interactions, the agent aims to learn a precoder policy that minimizes the BER performance. To this end, we develop the RL agent with DQN and DDPG that can find an optimal precoding policy in codebook-based and non-codebook based precoding systems, respectively. 
In order to demonstrate the optimality and robustness of proposed precoding framework, we explicitly consider two MIMO environments: {\it Environment I}, for which the optimal solution can be obtained by analytical approach, and {\it Environment II}, for which the optimal solution is not known. 
The following assumptions are made about MIMO systems and channel models in the two environments:
\begin{itemize}
\item {\bf{Environment I}} : a simple toy environment for which an optimal precoding vector is known. The toy environment consists of a MIMO-OFDM system with wideband precoding application and a flat-fading MIMO channel model. This toy scenario will be used to demonstrate the optimality of deep RL-based precoding framework. 
\item {\bf{Environment II}} : a realistic reference environment for which no optimal solution is known. The reference environment consists of a MIMO-OFDM system with subband precoding application and a frequency-selective MIMO channel model, which is representative of MIMO precoding applications in real-world deployments. To simulate over frequency-selective fading, we use a simple tap-delay-line channel model with two equal power taps and 400 ns tap spacing, i.e., the channel power profile [0, 0] dB with the tap delays [0,~400] ns. The channel model is chosen for easy learning and reproducibility in evaluating the proposed deep RL framework and conventional approaches. This reference scenario is used to demonstrate the robustness of deep RL-based precoding framework. 
\end{itemize}
For both environments, we assume that the channel follows a block-fading model, where the channel matrix on each subcarrier stays constant during a TTI but varies randomly from TTI to TTI.
Under such block-fading model, as illustrated in Fig. \ref{fig:prb}, only one of 14 OFDM symbols at each TTI is needed for transmitting the pilot signals to obtain an environmental state information. 
We note that in general RL problems modeled as MDP, the next state of the environment is determined as a function of the current state and the action taken by the agent. This implies that the agent will have to take into account the next state alongside the immediate reward when deciding which action to take. However, in our MIMO environment the next state is decided by a given channel model that only depends on the current state regardless of the action taken.
Since the agent's actions do not influence future states of the MIMO environment, the MIMO precoding problem can be regarded as a contextual bandit problem in which the agent's aim is to maximize an immediate reward at each time step. 
In other words, the goal of the DQN and DDPG agent is to choose a precoder $\vw$ from the pre-defined precoder space based on the channel matrices on the pilot channels that minimizes the immediate BER performance in (\ref{eq:bermin}). 

It is also worthy to note that while the precoder selection problem in Environment I can be easily solved by using an analytical solution under full knowledge of the underlying MIMO system and channel model, the task is a challenge for the RL agent due to two main reasons: First, the agent should learn the policy of choosing the best precoder solely based on the feedback of rewards without any knowledge on the underlying system and channel model. Second, finding an optimal policy in RL tasks with multidimensional continuous action spaces is known to be very difficult. 

% Grassmannian codebook
Generalization is an important problem in the action space design. The generalization over action spaces means that similar actions in similar states tend to have similar action values, which further imply that nearby states can have similar optimal actions. 
In the DQN, the action space will also play an important role in learning. We can formulate the action set design through a quantization process. That is, the action set can be obtained by quantizing the optimal action space under target environments. By assuming spatially-uncorrelated i.i.d. Rayleigh fading matrix channel model, the action set with the desired characteristics of quantization can be obtained by using a Grassmannian codebook proposed in \cite{Love:03}. 
In this paper, we utilize the Grassmannian codebook with size $N=64$ for codebook-based precoding.

% ------------------------------------------- %
\subsection{Optimality in Environment I}
In order to demonstrate the optimality of DRL-based precoding framework, 
we consider Environment I. 
We first provide the optimal solutions to codebook-based and non-codebook based precoding designs under Environment I and then describe how the DQN and DDPG algorithms can be applied for solving the same problems. We provide the simulation results to demonstrate that the DQN and DDPG-based agents can learn the near-optimal policies under Environment I.

% analytic optimal solution for Environment I
Under the assumption of the underlying wideband precoding MIMO system and flat-fading MIMO channel model in Environment I, the environmental state of MIMO channel can be fully captured by a channel matrix, denoted by $\mH_t$, on the single pilot signal in the reverse link at TTI $t$. In this case, the BER minimization problem in \eqref{eq:bermin} is reduced to a maximization problem of effective channel gain given by
\be 
g_{t}=|\vw^{\dag}\mH_t^{\dag}\mH_t\vw|.
\ee

As a result, in codebook-based MIMO precoding, the best precoder can be found by the following exhaustive search
\be \label{eq:optd}
\vw_{\text{d}}^{\text{opt}}=\argmax_{\vw \in \cA_d} |\vw^{\dag}\mH_t^{\dag}\mH_t\vw|,
\ee
where the discrete finite action space $\cA_d$ is given by the codebook.

Similarly, the optimal precoder in non-codebook based MIMO precoding mode is given by the following maximization problem
\be \label{eq:optc}
\vw_{\text{c}}^{\text{opt}}=\argmax_{\vw \in \cA_c} |\vw^{\dag}\mH_t^{\dag}\mH_t\vw|,
\ee
where the continuous action space $\cA_c$ corresponds to the surface of the unit sphere in $\dC^{n_{tx}}$ under the total power constraint described in Section \ref{sec:2}, i.e., $\cA_c=\left\{\vw | \vw \in \dC^{n_{tx}} s.t. \left|\left|\vw\right|\right|^2=1 \right\}$,
and the optimal solution $\vw_k^{\opt}$ is known to be given by the singular value decomposition (SVD) of $\mH_t$. 

The two optimal solutions in \eqref{eq:optd} and in \eqref{eq:optc} provide a strict lower–bound to the BER performance that can be achieved by the deep RL methods in codebook-based and non-codebook based MIMO precoding systems.
The two lower-bounds will be referred to as "{\it lower-bound to DQN}" and "{\it lower-bound to DDPG}", respectively.

% DQN to Environment I
\subsubsection{Codebook-based precoding.} 
We first elaborate on the application of DQN for estimating the state-action value function for each discrete precoder $\vw$ in a codebook $\cA_d$ through interactions with environments of the codebook-based MIMO system. 
As shown in Fig. \ref{fig:dqn}, the DQN $\bQ _{\theta}$ takes channel state $s$ as input and produces a distinct output for each action $a$ or precoder $\vw \in\cA_d$. 
As the state of a MIMO system is represented by a single channel matrix $\mH_t$, the channel state $s_t$ at TTI $t$ is given by a vector of size $2n_{tx}n_{rx}$ filled with the entries of $\mH_t$ as follows:
\be \label{eq:s}
s_t=o(\mH_t)=\left[\text{vec}(\Re\left[\mH_t\right])^{T}, \text{vec}(\Im\left[\mH_t\right])^{T}\right]^{T},
\ee
where $(\cdot)^{T}$ indicates the matrix transpose, $\text{vec}(\cdot)$ denotes the vectorization operator, while $\Re \left[\cdot \right]$ and $\Im \left[\cdot\right]$ denote the real and imaginary parts of a complex-valued argument. 

At each time step $t$, the DQN agent observes a context vector $s_t$ given by \eqref{eq:s} and chooses a precoder $\vw_t$ from the pre-defined codebook $\cA_d$ according to the $\epsilon$-greedy strategy in \eqref{eq:egreedy} to serve the MIMO system. After each time step, the agent receives a feedback of the experimental BER, in return for the action taken. Under the wideband precoding assumption, the experimental BER is obtained by
\be \label{ep:ber2}
\ber_t^{\ex}\left(\vw_t|\mH_t\right) = \frac{\sum_{ i=1}^{n_{RE}} \sum_{m = 1}^{\log_2M} \mathbbm{1}(b_i^m \ne \hat{b}_i^m)}{n_{RE}\log_2M}, 
\ee
where $n_{RE}$ is the total number of data REs used at each TTI and $\hat{b}_i^m$ denotes the hard decision of $b_i^m$ at the receiver and $\mathbbm{1}(\text{C})$ is an indicator that yields $1$ if C is true and $0$ if it is false. 

Since the experimental BER performance in \eqref{ep:ber2} represents the value of precoder $\vw_t$ in the channel state $s_t$ given by $\mH_t$ in \eqref{eq:s}, we can define the reward by the experimental BER performance. In particular, we use the following stochastic reward function 
\be \label{eq:rw}
r_t=\log_2\left(1-\ber_t^{\ex}\!\left(\vw_t|\mH_t\right)\right)+0.5,
\ee
where the logarithmic transformation of the bit-success rate $\left(1-\ber_t^{\ex}\!\left(\vw_t|\mH_t\right) \right)$ and a shift of $0.5$ are applied to produce the reward in the range of $[-0.5,0.5]$.

From the experience over the time steps, the agent learns about how the states $s_t$ and actions $a_t$ relate to each other so that the agent can predict the best precoder by observing the new state at the next step. In summary, a pseudo-code of the DQN algorithm with an $\epsilon$-greedy strategy is presented in Algorithm \ref{al:dqn}.

\begin{algorithm}[H]
\caption{DQN with decaying $\epsilon$-greedy}
\begin{algorithmic}[1]
\State{Initialize action-value network $\bQ_{\theta}$ with random parameters $\theta$ \Comment{using Xavier scheme \cite{Glorot:10} }} 
\State $\epsilon\gets 1$
\For{episode \texttt{ep=1,E}}
\State Initialize the initial state $s_0$ of sequence according to the channel model
\For{time step \texttt{t=0,T-1}}
\State{Choose an action $ \vw_t=\begin{cases}
\mbox{with probability } \epsilon, \mbox{select a random action from $ \cA_d$ } \\
\mbox{otherwise, take a greedy action by} \argmax_{a \in \cA_d} \bQ_{\theta}(s_t=o(\mH_t),a )
\end{cases}$}
\State{Execute action $\vw_t$ in environment and observe $\ber_t^{\ex}$} to get reward $r_t=\log_2\left(1-\ber_t^{\ex}\right)+0.5$
\State{Observe the next state $s_{t+1}=o(\mH_{t+1})$ according to the channel model}
\State{Compute the loss function $L({\theta})$ w.r.t target value $Y_t^{\theta}$ and on-line value $ \bQ _{\theta}(s_t,\vw_t)$ }
\State{Compute the gradient vector $\triangledown_{\theta} L({\theta})$ on the experience $\left[s_t,\vw_t,r_t,s_{t+1}\right]$ }
\State{Perform a gradient descent update w.r.t. $\theta$ as $ \theta \leftarrow \theta-\eta\triangledown_{\theta} L({\theta})$}
\EndFor
\State $\epsilon\gets \frac{1}{\texttt{ep}*5}$
\EndFor
\end{algorithmic} \label{al:dqn}
\end{algorithm}

Finally, the trained DQN can be used to choose the precoder by using the following arg-max operation
\be 
\vw_d^{\text{dqn}} =\argmax_{\vw \in \cA_d} \bQ_{\theta} (s, a=\vw).
\ee

% DDPG to Environment I
%% Figure DDPG
\begin{figure}%[ht]
\begin{subfigure}{.5\textwidth}
\centering
% include first image
\includegraphics[trim=0cm -0.5cm 0cm 0cm,clip,width=0.84\linewidth]{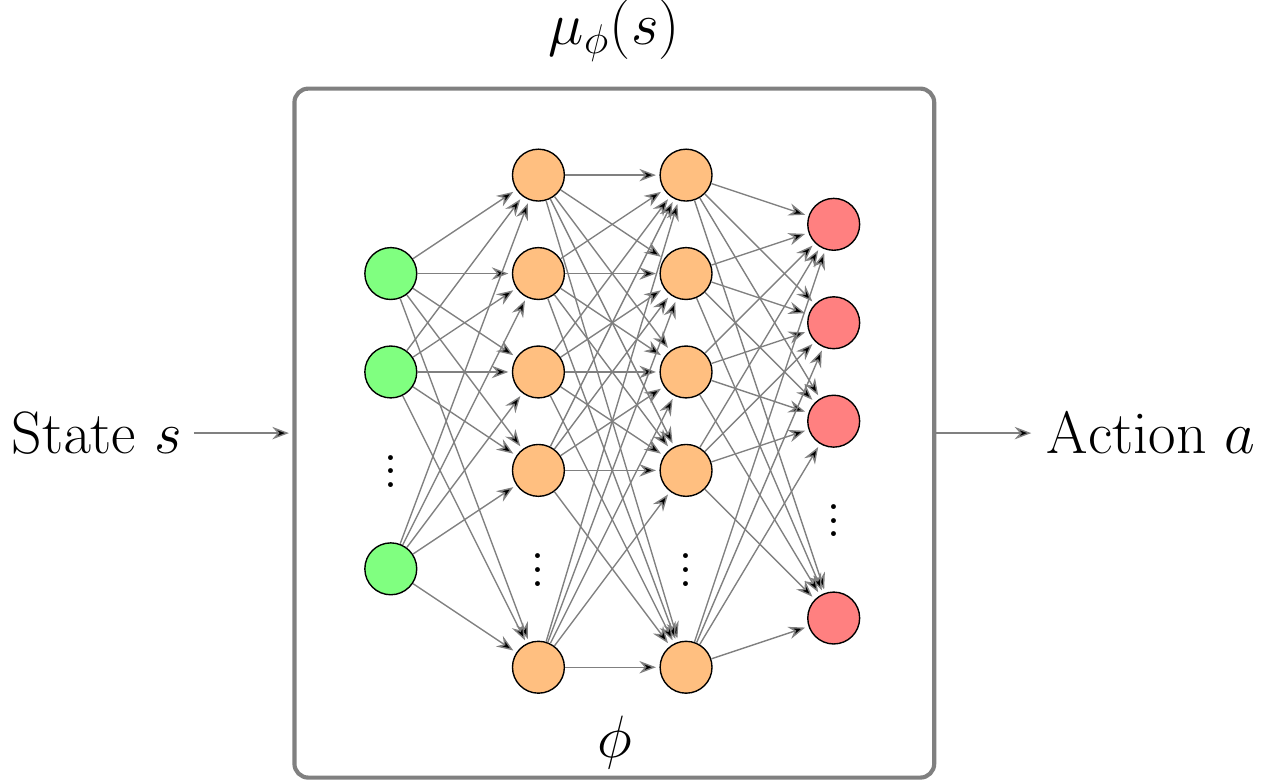}
\caption{actor with parameters $\phi$}
\label{fig:ddpg1}
\end{subfigure}
\begin{subfigure}{.5\textwidth}
\centering
% include second image
\includegraphics[trim=0cm -0.5cm 0cm 0cm,clip,width=1.0\linewidth]{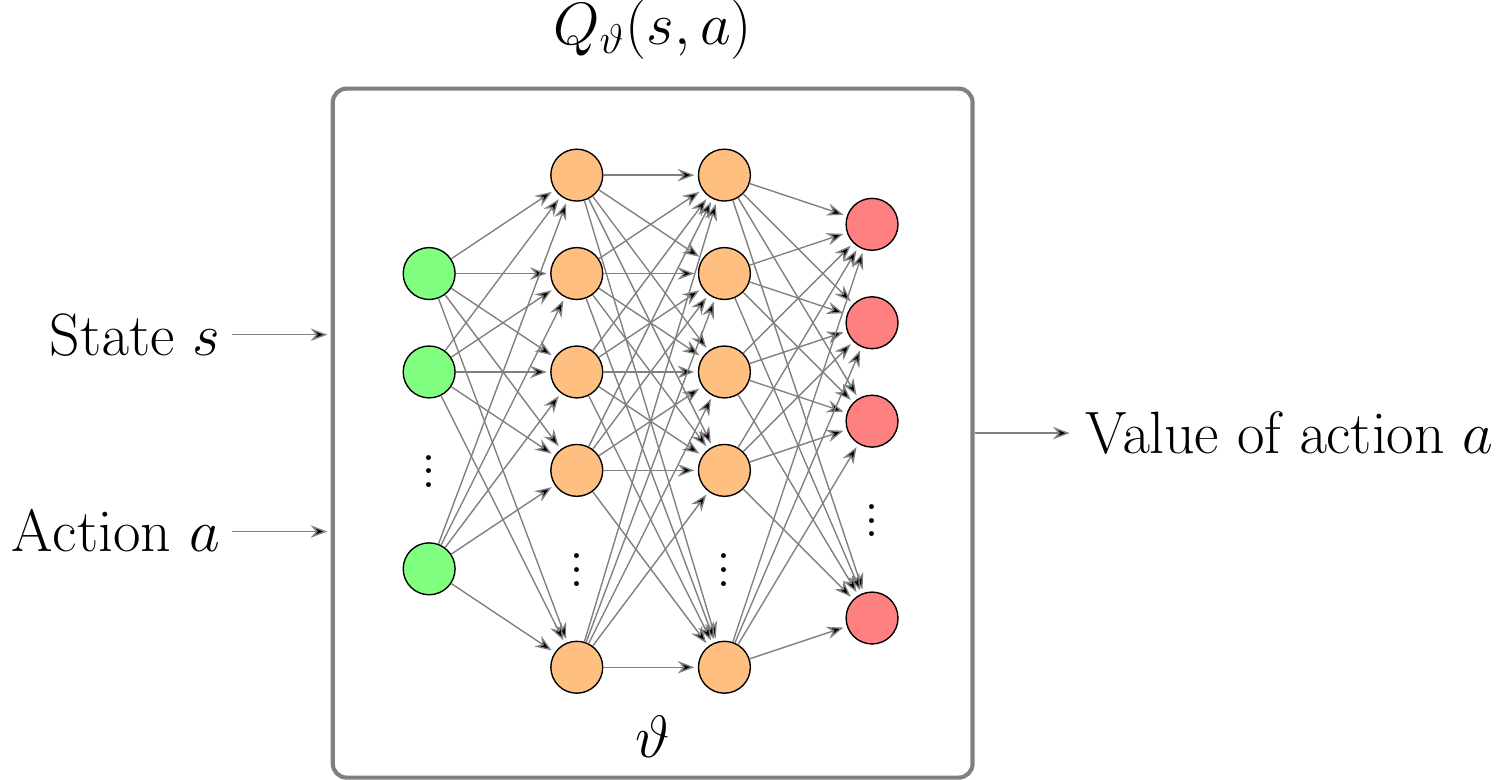}
\caption{critic with parameters $\vartheta$}
\label{fig:ddpg2}
\end{subfigure}
\caption{A schematic illustration of the policy-based DDPG that can be used for solving the non-codebook-based precoding problem in the proposed precoding framework, assuming a continuous action space $\cA_c$.}
\label{fig:ddpg}
\end{figure}
\subsubsection{Non-codebook-based precoding.} 
As noted above, the DQN cannot be applied the non-codebook based precoding problem because the arg-max operation in infinite action space becomes intractable. 
DDPG inherently provides an ability to learn an optimal precoding policy in the non-codebook based precoding mode. 
As described in Section \ref{sec:3}, DDPG concurrently learns a Q-function by the critic network $\bQ _{\vartheta}(s,a)$ and a policy by the actor network $\mu_{\phi}(s)$.
Figure \ref{fig:ddpg} illustrates the actor and critic networks of DDPG function approximator, where the actor takes state $s$ as input and provides a deterministic precoder $a$ (or $\vw$) in the continuous precoder space, and the critic takes not only state $s$ but also action $a$ as input and provides an action value of the given input as output. Note that unlike the DQN $\bQ _{\theta}(s,a)$ illustrated in Fig. \ref{fig:dqn} that takes only the state as input, the critic $\bQ _{\vartheta}(s,a)$ becomes able to deal with continuous action space by taking both action and state as input. 

The critic $\bQ _{\vartheta}(s,a)$ is trained with the Q-learning in the same way as the DQN described in Section \ref{sec:3} except the estimation of the target value, which is given by 
\be \label{eq:new2}
Y_t^{\vartheta}=r_t + \gamma \bQ _{\theta}(s_{t+1}, \mu_{\phi}(s_{t+1}) ),
\ee 
where compared to the DQN that assumes a greedy action on the next step in the evaluation of the target value, as shown in \eqref{eq:new}, the Q-value at the next state is evaluated by assuming the deterministic action $a=\mu_{\phi}(s_{t+1})$.

In the mean time, the actor network $\mu_{\phi}(s)$ is trained by utilizing the gradient of the critic $\bQ _{\vartheta}(s,a)$ with respect to action as follows:
\be
\triangledown_{\phi} J(\mu_{\phi}) = \triangledown_{\phi}\mu_{\phi}(s)|_{s=s_t} \triangledown_{a} \bQ _{\vartheta}(s,a)|_{s=s_t,a=\mu_{\phi}(s_t)}.
\ee

To ensure exploration during the training phase, the DRL algorithms define stochastic behavior policies in the training phase.
Figure \ref{fig:expl} compares different strategies for solving the exploration problems in the off-policy DQN and DDPG algorithms. As shown in Algorithm \ref{al:dqn}, the DQN algorithm uses an $\epsilon$-greedy strategy in a discrete action space by selecting a random action with a certain probability. In order to perform exploration in continuous action spaces, DDPG perturbs the action chosen by the deterministic policy by adding a noise vector, e.g., 
\be
a=\mu_{\phi}(s)+\nu,
\ee
where $\nu \in \dR^{2n_{tx}}$ (or $ \nu \in \dC^{n_{tx}}$) is an additive white Gaussian noise (AWGN) vector whose elements are independent identically distributed (i.i.d.) complex-valued Gaussians with zero mean and variance $\sigma_p^2$.

Based on the deterministic gradient, the DDPG can solve complex tasks with high-dimensional continuous action spaces. 
A pseudo-code of the DDPG application to the non-codebook based precoding design is presented in Algorithm \ref{al:ddpg}.

\begin{algorithm}[H]
\caption{DDPG}
\begin{algorithmic}[1]
\State{Initialize actor network $\mu_{\phi}(s)$ and critic network $\bQ _{\vartheta}(s,a)$ with random parameters $\phi$ and $\vartheta$\Comment{using Xavier scheme \cite{Glorot:10} }} 
\State $\epsilon\gets 1$
\For{episode \texttt{ep=1,E}}
\State Initialize the initial state $s_0$ of sequence according to the channel model
\For{time step \texttt{t=0,T-1}}
\State{Choose an action $ a_t=\mu_{\phi}(s_t=o(\mH_t)) + \nu$ with exploration noise $\nu$}
\State{Execute action $a_t$ in environment and observe $\ber_t^{\ex}$} to get reward $r_t=\log_2\left(1-\ber_t^{\ex}\right)+0.5$
\State{Observe the next state $s_{t+1}=o(\mH_{t+1})$ according to the channel model}
\State{Compute the loss function $L({\vartheta})$ w.r.t target value $Y_t^{\vartheta}$ and on-line value $ \bQ _{\vartheta}(s_t,a_t)$ }
\State{Compute the gradient vector $\triangledown_{\theta} L({\vartheta})$ on the experience $\left[s_t,a_t,r_t,s_{t+1}\right]$ }
\State{Update the critic network $\vartheta$ as $\vartheta \leftarrow \vartheta-\eta\triangledown_{\vartheta} L({\vartheta})$}
\State{Compute the gradient vector $\triangledown_{\phi} J(\mu_{\phi}) = \triangledown_{\phi}\mu_{\phi}(s)|_{s=s_t} \triangledown_{a} \bQ _{\vartheta}(s,a)|_{s=s_t,a=\mu_{\phi}(s_t)}$ }
\State{Update the actor network $\phi$ as $\phi \leftarrow \phi+\eta \triangledown_{\phi} J(\mu_{\phi}) $}
\EndFor
\EndFor
\end{algorithmic} \label{al:ddpg}
\end{algorithm}

Then, in the execution phase, the learned actor $ \mu_{\phi}$ is used to choose the precoder 
\be 
\vw_c^{\text{ddpg}} = \mu_{\phi}(s).
\ee

% Simulation results
Finally, we present simulation results to compare the DRL-based solutions with the two lower bounds in \eqref{eq:optd} and \eqref{eq:optc}. We evaluate BER performance over 4-by-2 flat-fading MIMO-OFDM system consisting of 960 subcarriers with $30$ kHz subcarrier spacing and using the 16-QAM modulation. 
We implemented the DQN and DDPG algorithms in TensorFlow using a fully connected neural network with two hidden layers that have 512 and 128 neurons, respectively, and use the rectified linear unit (ReLU) as the activation functions. For fast and stable convergence, we initialize the weights randomly from a normal distribution by using the Xavier scheme \cite{Glorot:10} while the biases are initialized to be zero. 

%% Figure cdf
\begin{figure}%[htbp]
\centerline{\includegraphics[trim=0cm 0cm 0cm 0cm,clip,width=0.55\linewidth]{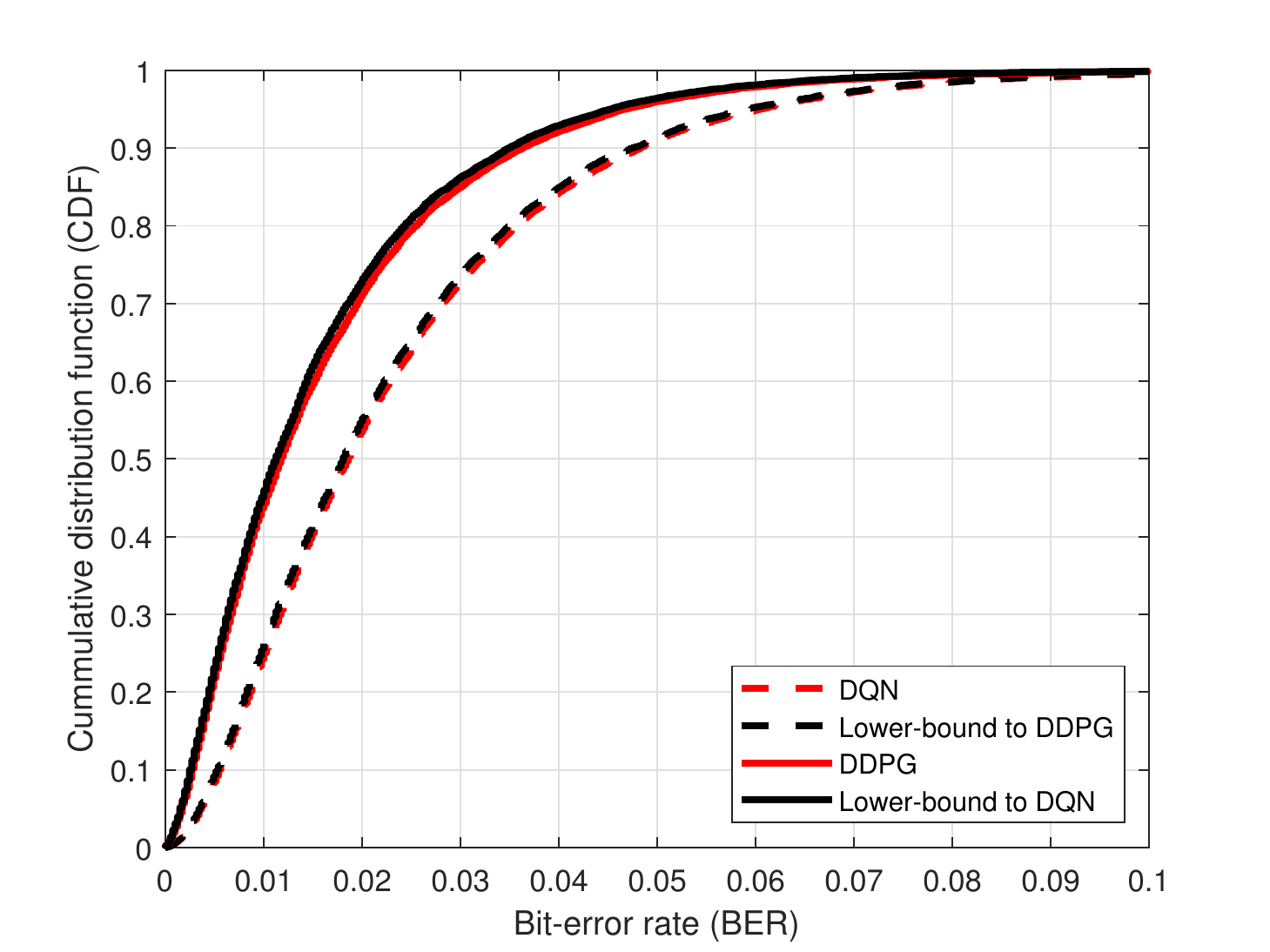}}
\caption{Cumulative distribution function (CDF) of the bit-error rate (BER)-based rewards of the 4-by-2 MIMO OFDM system over the flat-fading channel model using 16-QAM}
\label{fig:cdf}
\end{figure}
The achieved BER performance of the DQN and DDPG is presented in Fig. \ref{fig:cdf} in comparison with the two lower bounds. 
The DQN and DDPG are learned over 300,000 time steps under the temporally-independent block-fading channel model and the performance is measured on a new set of 10,000 steps without parameter update. 
The comparison with the lower bounds shows that the DQN and DDPG can achieve the near-optimal performance of codebook-based and non-codebook based MIMO precoding system.
The results demonstrate that the DQN and DDPG are able to learn the near-optimal precoder selection policy solely based on the feedback of rewards without any additional knowledge on the underlying system and channel model in the wireless communication environment. 

% ------------------------------------------- %
\subsection{Robustness in Environment II}
Motivated by the optimality of DRL-based precoding approach demonstrated in Environment I, 
we aim at exploiting the benefit of the approach on a more challenging precoding problem under Environment II. 
We first provide the sub-optimal analytic solutions to the precoding problems in codebook-based and non-codebook based MIMO transmissions under Environment II and describe the DQN and DDPG applications to the same problems. We then provide the simulation results to demonstrate the robustness of the two DRL algorithms to learn an optimal policy under very complex environments.

% sub solution for environment II
As shown in Section \ref{sec:2}, under the assumption of the underlying subband precoding MIMO system and frequency-selective MIMO channel model in Environment II, the BER minimization problem in \eqref{eq:bermin} does not admit a computationally efficient solution. A conventional approach is to use the spatial channel statistics of the pilot channels within a subband. 
Under the block-fading model, the environmental state (or the spatial channel statistics) can be captured by channel matrices $\{\mH_{t,j}\}_{j\in \Psi_{p}}$ estimated on the reverse pilot signals. 
The spatial channel statistics can be approximated by the channel covariance matrix averaged over the pilot channels within a subband, i.e.,
\be \label{eq:evd}
\mR_{t,hh}=\frac{1}{|\Psi_{p}|} \sum_{j \in \Psi_{p}} \mH_{t,j}^{\dagger} \; \mH_{t,j}.
\ee

By utilizing this spatial channel covariance matrix for the conventional channel gain maximization problems in \eqref{eq:optd} and \eqref{eq:optc}, the analytical suboptimal solutions can be computed. 

Accordingly, in codebook-based MIMO precoding, the best precoder is given by the following exhaustive search:
\be \label{eq:subd}
\vw_{\text{d}}^{\text{sub-opt}}=\argmax_{\vw \in \cA_d} |\vw^{\dag}\mR_{t,hh}\vw|,
\ee
and the best sub-optimal precoder in non-codebook based MIMO precoding is given by the following maximization problem:
\be \label{eq:subc}
\vw_{\text{c}}^{\text{sub-opt}}=\argmax_{\vw \in \cA_c} |\vw^{\dag}\mR_{t,hh}\vw|,
\ee
where the sub-optimal solution $\vw_{\text{c}}^{\text{sub-opt}}$ is known to be obtained by the eigenvalue decomposition (EVD) of $\mR_{t,hh}$ \cite{Telatar:99}. 

% DRL to environment II
Nevertheless, the above conventional approximation solutions are far from being optimal due to the approximation steps taken to simplify the BER minimization problem in \eqref{eq:bermin} into \eqref{eq:subd} or \eqref{eq:subc}. In what follows, we consider a DRL-based approach as an alternative. That is, instead of approximating an optimal precoder based on the spatial channel covariance matrix, the proposed DRL-based framework learns an optimal precoding policy directly from interactions with complex real-world MIMO environments. The DRL-based approach will lead to a solution that is closer to the optimum for the original precoding problems \eqref{eq:bermin} with $\cA=\cA_d$ or $\cA_c$.

The same DQN and DDPG algorithms illustrated in Algorithm \ref{al:dqn} and \ref{al:ddpg} can be used under Environment II while the input of environmental state to the neural networks is a three-dimensional array representing the transmit antenna, the receive antenna, and the RE in a given subband. As we use a fully-connected input layer in our simulations, the environmental state vector $s_t$ for a given subband is given by a set of vectorized MIMO channel matrices on the pilot REs $\Psi_{p}$
\be
s_t=o(\{\mH_{t,j}\}_{j\in \Phi_{d}})=\left\{\left[\text{vec}(\Re\left[\mH_{t,j}\right])^{T}, \text{vec}(\Im\left[\mH_{t,j}\right])^{T}\right]^{T}\right\}_{j\in \Psi_{p}},
\ee
and, under the subband precoding assumption, the experimental BER over the data REs $ \Phi_{d}$ is given by 
\be 
\ber_t^{\ex}\left(\vw_t |\{\mH_{t,j}\}_{j\in \Phi_{d}}\right) \triangleq \frac{1}{\log_2M|\Phi_{d}|} \sum_{m=1}^{\log_2M} \sum_{j=1}^{|\Phi_{d}|}\mathbbm{1}(b_i^m \ne \hat{b}_i^m).
\ee

% Simulation results
To demonstrate the robustness of the DQN and DDPG in learning an optimal solution under Environment II, we provide numerical results, comparing those with the conventional approximation algorithms in \eqref{eq:subd} and \eqref{eq:subc} under the frequency-selective TDL channel model. We consider the same 4-by-2 MIMO-OFDM system setup used for Environment I, but here we also evaluate the 4-QAM modulation. The subband size is assumed to be 8 PRBs. In favor of the conventional solution, the number of pilot signals per subband is chosen to be 3, beyond which only a marginal gain was observed for the conventional solution. 

Here the DQN and DDPG both have three hidden fully-connected layers. 
The three hidden layers have 3840, 512 and 128 neurons, respectively, and use the rectified linear unit (ReLU) as the activation functions. 
It is worth mentioning that here we did not aim at optimizing the neural network in terms of reduced numbers of layers and neurons for our agent because there exist network compression methods, such as weight pruning and quantization, that can dramatically reduce the computation and memory requirements without affecting the learning performance \cite{Han:16}. Moreover, new gradient-based optimization methods have been proposed that improve the basic stochastic gradient descent algorithm, including the adaptive learning rate and gradient updates \cite{Kingma:15}, \cite{Reddi:18}.

%% Figure cdf2
\begin{figure}%[htbp]
\centerline{\includegraphics[trim=0cm 0cm 0cm 0cm,clip,width=0.55\linewidth]{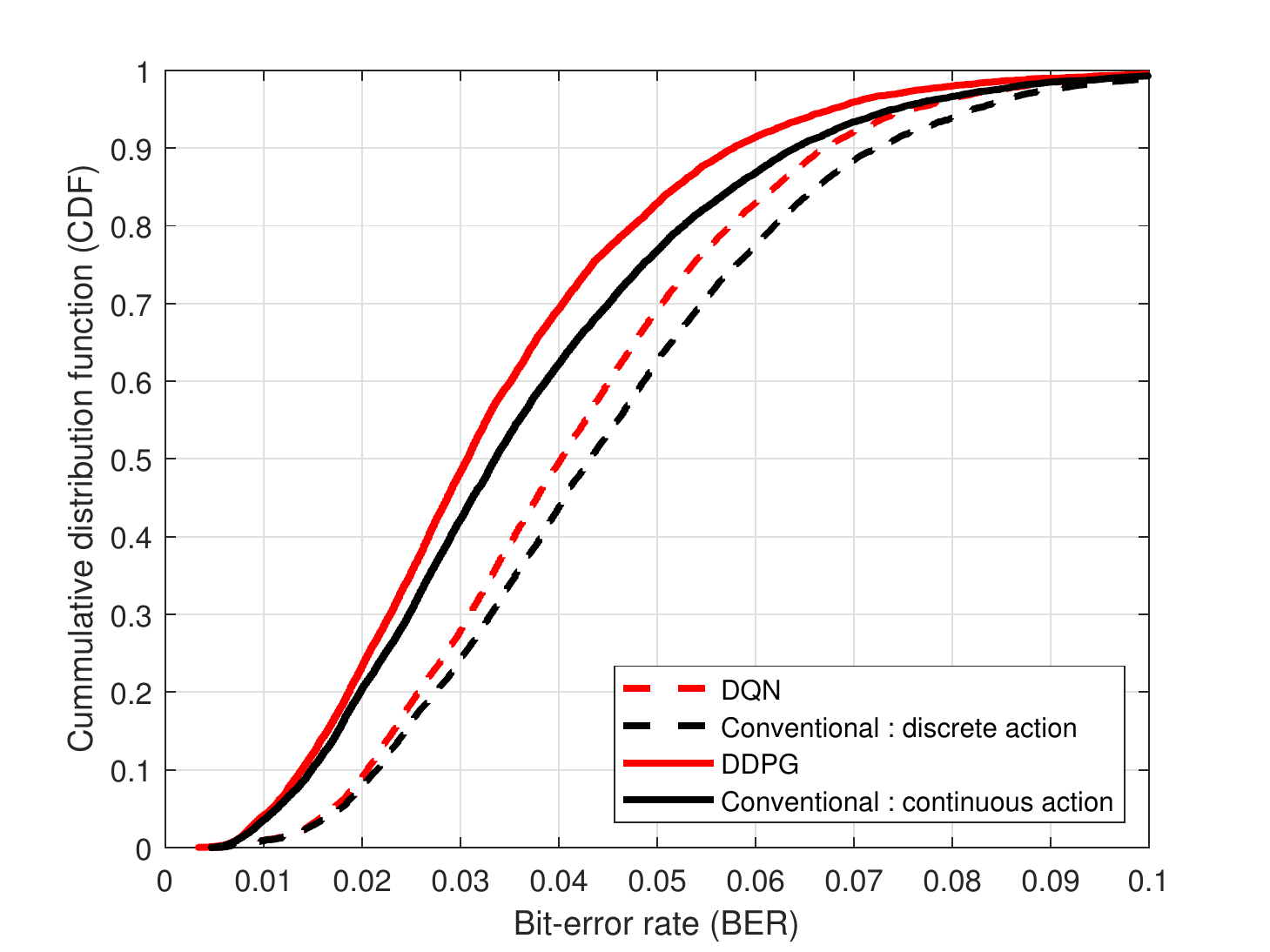}}
\caption{Cumulative distribution function (CDF) of the bit-error rate (BER)-based rewards of the 4-by-2 MIMO OFDM system over the two-tap TDL channel model using 16-QAM}
\label{fig:cdf2}
\end{figure}

%% Figure BER
\begin{figure}%[htbp]
\centerline{\includegraphics[trim=0cm 0cm 0cm 0cm,clip,width=0.55\linewidth]{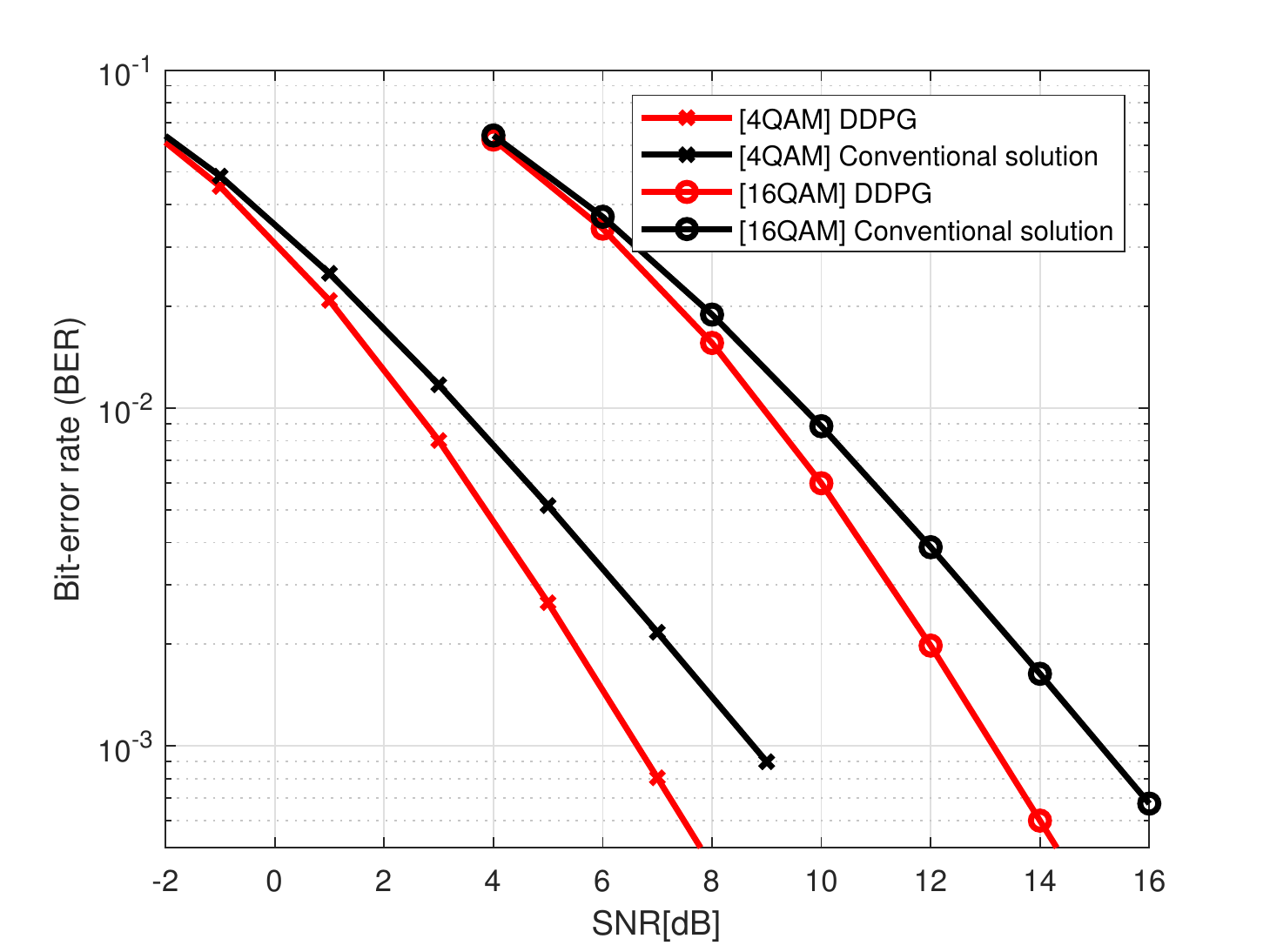}}
\caption{Bit-error rate (BER) of the 4-by-2 MIMO OFDM system over the two-tap TDL channel model using 4-QAM and 16-QAM}
\label{fig:ber}
\end{figure}

The DQN and DDPG are pretrained over 3,000,000 subbands and the performance is measured on a new set of 10,000 subband without parameter update. 
The achieved BER performance of the proposed precoding framework is presented in the figures \ref{fig:cdf2} and \ref{fig:ber} in comparison with the conventional approximation solutions in \eqref{eq:subd} and \eqref{eq:subc}. 
The simulation results show that the proposed precoding framework is able to learn a good precoding policy under the very complex environment, outperforming the conventional algorithms in both codebook-based and non-codebook based MIMO precoding systems.

% ------------------------------------------------------------------------------ Conclusion ------------------------------------------------------------------------------%
\section{Conclusion} \label{sec:5}
In this paper, we have proposed a DRL-based precoding framework that can be used to learn an optimal
precoding policy for complex MIMO precoding problems. In particular, we applied two leading DRL algorithms, deep Q-network (DQN) and deep deterministic policy gradient (DDPG), to codebook-based and non-codebook based precoding problems and showed that there is a natural fit between the ideas of DQN and DDPG algorithms and the principles of codebook-based and non-codebook based precoding modes. We have shown the optimality and robustness of the proposed precoding framework by comparing its performance with that of the conventional optimal solution in a simple MIMO environment and the best sub-optimal solution in a complex MIMO environment in terms of achieved bit-error rate (BER). Specifically, the simulation results have demonstrated that the DRL-based approach has the potential to outperform the existing algorithms in complex wireless communication environments for which no optimal solutions are known.
Based on our results, we believe that the proposed DRL framework will offer a promising physical-layer solution for future wireless systems. 

\bibliographystyle{ieeetr}
\bibliography{REFbib}

\end{document}